\documentclass[acmtog]{acmart}

\acmSubmissionID{543}

\usepackage{booktabs} %

\usepackage[ruled]{algorithm2e} %

\SetAlFnt{\small}
\SetAlCapFnt{\small}
\SetAlCapNameFnt{\small}
\SetAlCapHSkip{0pt}

\acmJournal{TOG}
\acmVolume{42}
\acmNumber{6}
\acmArticle{112.1522}
\acmYear{2023}
\acmMonth{12}

\acmDOI{10.1145/3618375}

\usepackage{subcaption}
\usepackage{soul}
\usepackage{multirow}
\usepackage{multicol}
\usepackage{array}

\begin{document}
\title{AdaptNet: Policy Adaptation for Physics-Based Character Control}

\author{Pei Xu}
\orcid{0000-0001-7851-3971}
\affiliation{%
 \institution{Clemson University}
 \country{USA}}
 \affiliation{
 \institution{Roblox}
 \country{USA}}
\email{peix@clemson.edu}

\author{Kaixiang Xie}
\orcid{0000-0002-5877-9374}
\affiliation{%
 \institution{McGill University}
 \country{Canada}}
\email{kaixiang.xie@mail.mcgill.ca}

\author{Sheldon Andrews}
\orcid{0000-0001-9776-117X}
\affiliation{%
 \institution{École de Technologie Supérieure}
 \country{Canada}}
 \affiliation{
 \institution{Roblox}
 \country{USA}}
\email{sheldon.andrews@gmail.com}

\author{Paul G. Kry}
\orcid{0000-0003-4176-6857}
\affiliation{%
 \institution{McGill University}
 \country{Canada}}
\email{kry@cs.mcgill.ca}

\author{Michael Neff}
\orcid{0000-0003-0226-2808}
\affiliation{%
 \institution{University of California, Davis}
 \country{USA}}
\email{mpneff@ucdavis.edu}

\author{Morgan McGuire}
\orcid{0000-0003-1074-0953}
 \affiliation{
 \institution{Roblox}
 \country{USA}}
\affiliation{%
 \institution{University of Waterloo}
 \country{Canada}}
\email{morgan@roblox.com}

\author{Ioannis Karamouzas}
\orcid{0009-0000-4315-6556}
\affiliation{%
 \institution{University of California, Riverside}
 \country{USA}} 
\email{ioannis@cs.ucr.edu}

\author{Victor Zordan}
\orcid{0000-0002-7309-7013}
\affiliation{
 \institution{Roblox}
 \country{USA}}
\affiliation{%
 \institution{Clemson University}
 \country{USA}}
\email{vbzordan@roblox.com}

\begin{CCSXML}
<ccs2012>
   <concept>
       <concept_id>10010147.10010371.10010352</concept_id>
       <concept_desc>Computing methodologies~Animation</concept_desc>
       <concept_significance>500</concept_significance>
       </concept>
   <concept>
       <concept_id>10010147.10010371.10010352.10010379</concept_id>
       <concept_desc>Computing methodologies~Physical simulation</concept_desc>
       <concept_significance>300</concept_significance>
       </concept>
   <concept>
       <concept_id>10010147.10010257.10010258.10010261</concept_id>
       <concept_desc>Computing methodologies~Reinforcement learning</concept_desc>
       <concept_significance>300</concept_significance>
       </concept>
 </ccs2012>
\end{CCSXML}

\ccsdesc[500]{Computing methodologies~Animation}
\ccsdesc[300]{Computing methodologies~Physical simulation}
\ccsdesc[300]{Computing methodologies~Reinforcement learning}

\keywords{character animation, physics-based control, motion synthesis, reinforcement learning, motion style transfer, domain adaptation, GAN}

\renewcommand{\shortauthors}{P. Xu, K. Xie, S. Andrews, P. Kry, M. Neff, M. McGuire, I. Karamouzas, and V. Zordan}

\begin{abstract}
Motivated by humans' ability to adapt skills in the learning of new ones, this paper presents
AdaptNet, an approach 
for modifying 
the latent space of %
existing policies 
to allow  
new behaviors to be quickly learned from like tasks in comparison to learning from scratch.  
Building on top of a given reinforcement learning  controller, %
AdaptNet %
uses a two-tier hierarchy that augments the original state embedding to support modest changes in a behavior and further modifies the policy network %
layers to make more substantive changes.   
The technique is shown to be effective for adapting existing physics-based controllers to a wide range of new styles for locomotion, new task targets, changes in character morphology and extensive changes in environment. Furthermore, it exhibits
significant increase in learning efficiency, as indicated by greatly reduced training times when compared to training from scratch or using other approaches that modify existing policies.
Code is available at 
\href{https://motion-lab.github.io/AdaptNet}{\textit{https://motion-lab.github.io/AdaptNet}}.

\end{abstract}

\begin{teaserfigure}
\centering
    \includegraphics[width=\linewidth]{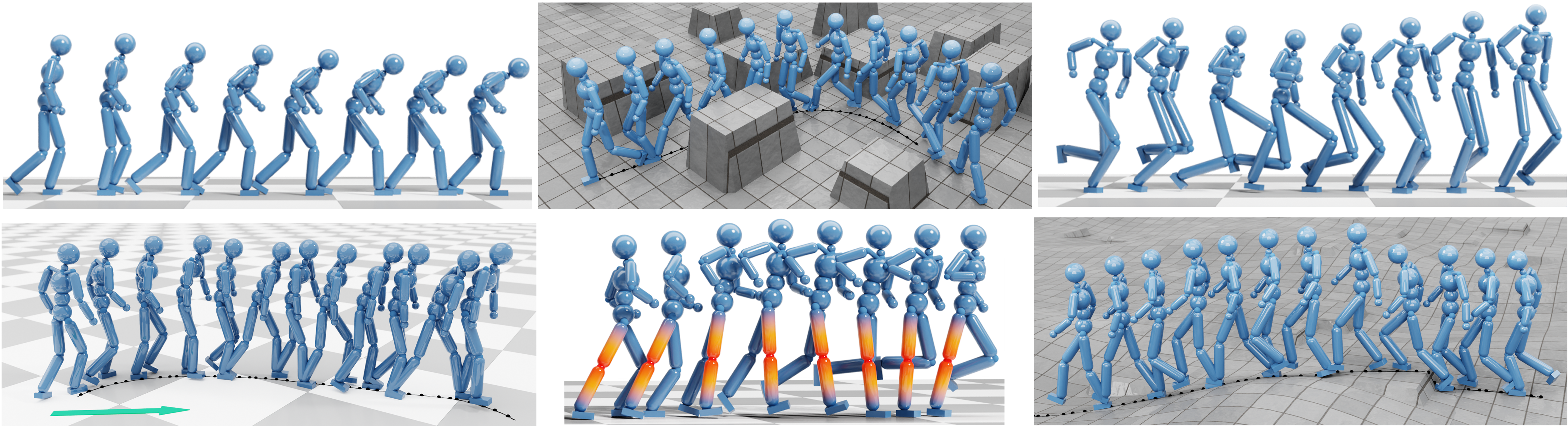}\hfill
    \caption{
    Examples policy adaptation for locomotion.
    From left to right and top to bottom: 
    motion interpolation, local collision avoidance, 
    body-length changes, style transfer, morphology changes, 
    rough terrain adaptation.  
}
    \label{fig:teaser}
\end{teaserfigure}

\maketitle

\section{Introduction}
Research on physically-based character animation has received a great deal of attention recently, especially using reinforcement learning (RL) to develop control policies that produce a wide spectrum of motion behaviors and styles with few or no manual inputs. 
Most techniques
rely on reference human motion to either provide direct tracking or indirect comparison to constrain movement, along with additional
targets and rewards to shape task success (e.g., \cite{Peng2018,Liu2018,iccgan}).
However, 
methods to date largely develop policies or controllers for a known behavior, and must be relearned (usually from scratch) to produce a new behavior.  While curriculum-style learning and warm-start approaches may be used to migrate policies to targeted goal tasks \cite{Yin2021,Tao2022},
we instead aim to broadly adapt previously trained policies to make them usable in a wide spectrum of new scenarios without the need for full retraining.

Inspired by recent work  %
in conditioning existing models in image-based stable diffusion and large language models \cite{zhang2023adding,hu2021lora}, we introduce \emph{AdaptNet} as an approach for controlling physically based characters that modifies an existing policy to produce behavior in a variety of new settings. 
The main novelty of our work is the ability to control the motion generation process through editing the latent space. 
In physics-based character control tasks, there is an opportunity to better understand and exploit the latent space representation of control policies obtained using reinforcement learning frameworks. AdaptNet provides an initial step in this direction.

Specifically, our approach relies on the training of weights for new network components that are injected into a previously trained policy network.  Building on top of a pre-existing multi-objective reinforcement learning %
controller, we propose a two-tier architecture for AdaptNet that augments the latent state 
embedding while adding modifications to the remaining layers for control refinement. 
The first layer modifies the latent space projected from the association of the task and character state.  It supports adding elements to the control state, as well as changing the imitation and task rewards.  Meanwhile, the deeper, control-level refinement augments the policy's action derived from the latent state, supporting more substantive changes to the task control.
Together, AdaptNet performs fast training from a previously trained policy and is capable of making a wide spectrum of adaptations from a single behavior.

As in Figure~\ref{fig:teaser}, we showcase our learning framework with numerous controller adaptation examples, including changes in the style of locomotion derived from very short reference motions.  AdaptNet can perform this ``few-shot style transfer'' using only the embedding layer augmentation in a fraction of the time it takes to learn the original locomotion policy.  Furthermore, through interpolating in the latent space, 
it is possible to control the generated control dynamically and smoothly transition from the original behavior to the new style. 
 We further experiment with changes to the character morphology by ``locking'' joints and changing limb lengths.  While these changes lead to failure in the original policy, AdaptNet augments the policy easily to account for the various changes.  We also investigate changes in the environment, exploring adaptation for locomotion on rough and slick (low-friction) terrains, as well as on obstacle-filled environments.  In each case, AdaptNet provides significant 
improvement leading to characters that robustly traverse a range of new settings (see Figure 1 and accompanying video).

We evaluate the effectiveness of AdaptNet on various tasks, %
including its ability for adaptation of imitation learning, different goal rewards, and environmental states.  We compare our approach %
against training from scratch, as well as training-continuation (finetuning). 
Training with AdaptNet can typically be carried out within 10-30 minutes for simple adaptation tasks, and up to 4 hours for complex locomotion tasks and environment changes.
Within such modest training time budgets, in most cases it is impossible to obtain a working controller that can adhere to imitation and goal-task objectives when training from scratch or finetuning a pre-existing policy. 
Additional ablation studies support the specific architecture we propose over several alternatives along with highlighting AdaptNet's ability to successfully control and modify the latent space.

The contributions of our work are summarized as follows:
\begin{itemize}

\item We show how the latent space representation of an RL policy can be modified for motion
synthesis in physics-based motor control tasks.  

\item  Based on this, we introduce AdaptNet as a framework to efficiently modify a pre-trained physics-based character controller to new tasks. 

\item  We showcase the applicability of AdaptNet on a variety of multi-objective adaptation tasks, including few-shot motion style transfer, motion interpolation, character morphology adaptation, and terrain adaptation.

\end{itemize}

\section{Related Work}
Our approach follows a wide set of previous related work stemming from general disciplines in computer animation, robotics, machine learning and image generation.  We focus on the background
that is most relevant, categorized in physically based character skill control, transfer learning, and latent space adaptation.

\subsection{Deep Reinforcement Learning for Skilled Motion}

Deep learning neural network control policies have become the staple 
for physics-based character animation research due to their ability to synthesize a range of skilled motions. In recent years, techniques have trained control policies to animate physics-based humanoid characters for agile motions~\cite{Yin2021}, team sports~\cite{Xie2022, Liu2018}, martial arts~\cite{won2021control}, juggling~\cite{Chemin2018,Luo2021,composite}, performing complex environment interactions~\cite{Merel2020}, as well as general locomotion tasks~\cite{Bergamin2019,Peng2018}. The recent survey by \citet{DeepSurvey2021} provides a comprehensive overview of approaches that have been developed for motion synthesis and control of animated characters. 

Training skill-specific policies often requires extended training time, necessitating years of simulated learning~\cite{peng2022ase}. Skill re-use and combining pre-trained policies to perform more complex tasks offer an alternative that can create needed savings from this extensive training. To this end, a number of papers have proposed ways to reuse and/or combine policies.  For example, DeepMimic~\cite{Peng2018} trains a composite policy that transitions between a collection of different skills. \citet{Liu2017} experiment with hierarchical models that sequence a set of pre-trained control fragments. \citet{hejna20a} explore a hierarchical approach to decouple low and high-level policies to transfer skills from agents with simple morphologies to more complex ones, and found that it helps to reduce overall sampling.  
Likewise, we demonstrate that the proposed AdaptNet approach is effective when adapting pre-trained policies to new character morphologies and motion styles with relatively little additional training time. 

Curriculum learning is also related to skill adaptation since the agent is trained on tasks with increasing difficulty~\cite{karpathy2012curriculum,symmetric}. 
The approach is demonstrated to be effective for training controllers that allow agents to traverse environments of increasing complexity~\cite{ALLSTEPS2020,heess2017emergence} and recover to standing~\cite{Frezzato2022} under increasingly challenging conditions. In comparison, we demonstrate that our approach efficiently allows a physically simulated humanoid to adapt pre-trained walking and running skills to new terrains as well. However, the aim for curriculum learning is somewhat different than our own in that it is usually used as a means to develop a single advanced skill while we focus on the ability to generalize from one behavior to many.  %

\subsection{Transfer Learning}
In machine learning, a common approach for model adaptation is to start with a pre-trained model and fine tune it on a new task. 
Over the years a number of architectures %
have been proposed to overcome %
the overfitting and expressivity issues of finetuning, 
including GAN-inspired approaches for domain adaptation~\cite{ganin2016domain,tzeng17} and 
adding new models to previously learnt ones through lateral connections~\cite{rusu2016progressive,rusu2017sim}.
To %
facilitate better model transfer, algorithms have been explored that account for entropy optimization~\cite{haarnoja2017reinforcement,wang2020tent}. As well, others directly manipulate the source task domain through randomizing physical parameters of the agent and/or environment while adapting the source domain to the target one~\cite{rajeswaran2017epopt,peng2018sim,ganin2016domain}. 
To encourage diversity during early training, recent work on transfer learning has also explored a multi-task paradigm where a model is pre-trained on many tasks before being transferred to a new target domain~\cite{devin2017learning,alet2018modular}.
Some multi-task transfer learning solutions include  
policy distillation that 
seeks to ``distill'' knowledge from expert policies to a target policy~\cite{rusu2015policy,ParisottoBS15}.   
Another approach with a similar goal is policy learning which learns a residual around given expert policies~\cite{silver2018residual}. 

Meta learning has also gained popularity recently in computer vision and robotics, seeking to leverage past experiences obtained from many tasks to acquire a more generalizable and faster model that can be quickly adapted to new tasks~\cite{andrychowicz2016learning,ravi2017optimization}. 
The related formulations can be broadly classified into models that ingest a history of past experiences through recurrent architectures~\cite{duan2016rl,heess2015memory}, model-agnostic meta-learning methods~\cite{finn2017model,nichol2018first}, and approaches for meta-learning hyperparameters, loss functions, and task-dependent exploration strategies~\cite{xu2018meta,houthooft2018evolved,gupta2018meta}. 

While some of the aforementioned approaches have shown great promise for agent control problems, 
in this paper, we propose an approach that can quickly adapt RL policies for physically simulated humanoids through fine control tuning as well as augmentation injected in the latent space, loosely inspired by recent findings in image diffusion~\cite{zhang2023adding,hu2021lora,mou2023t2iadapter}.
In character animation, 
related work has focused on motion style transfer tasks for \emph{kinematic} characters~\cite{mason2017,aberman2020} %
and the recent work of~\citet{starke2022deepphase} shows exciting results about how a well-learned latent space can aid motion synthesis. %
However, in physics-based character control tasks,
there is still little investigation about the latent space representation of the control policy obtained  using reinforcement learning frameworks. We believe that AdaptNet provides a promising step in bridging that gap.

\subsection{Latent Space Adaptation}

We are inspired by research in image and 3D model generation that shows it is possible to control the synthesis process to generate targeted artifacts through purposeful modification of the 
latent space%
~\cite{radford2015unsupervised,zhuang2021enjoy,epstein2022blobgan,pmlr-v80-bojanowski18a,abdal2019image2stylegan,shen2020interpreting,wu2016learning,berthelot2017began,karras2020analyzing}.
While we have seen related work in RL for character control, AdaptNet offers a unique approach to
latent space adaptation, drawn from these adjacent works' successes.
Related works in physics-based character control, such as~\cite{ling2020character,peng2022ase,2022-SA-PADL,tessler2023calm,won2021control, MCPPeng19}, explore using pre-trained latent space models to facilitate the training of a control policy. 
These methods intend to adapt the pre-trained multi-skill model for downstream tasks by controlling skill latent embeddings, focusing on reusing skills for motion generation.  
In contrast, 
our approach does not break down the latent space by task and character state and instead allows the policy to be adapted to heterogeneous tasks that require learning new (out-of-distribution) motions/skills.
Further, previous methods discard the pre-trained latent encoder during adaptation and rely on re-training to obtain a new encoder.
In contrast, 
our approach directly edits the latent space projected from the association of the task and character state via the pre-trained policy.  To do this, we use a
gated recurrent unit (GRU)~\cite{chung2014empirical} 
layer as the encoder and initialize it by duplicating the original encoder parameters.  Next, a fully connected layer is applied after the GRU to ensure zero initialization and convert the encoded state to a latent \emph{modification}. 
In sum, the training for our adaptation starts from modifying the pre-trained policy rather than from scratch, 
which benefits adaptation in comparison to previous work in sample efficiency and, at times, overall effectiveness.

\begin{figure}[t]
    \centering
    \includegraphics[width=\linewidth]{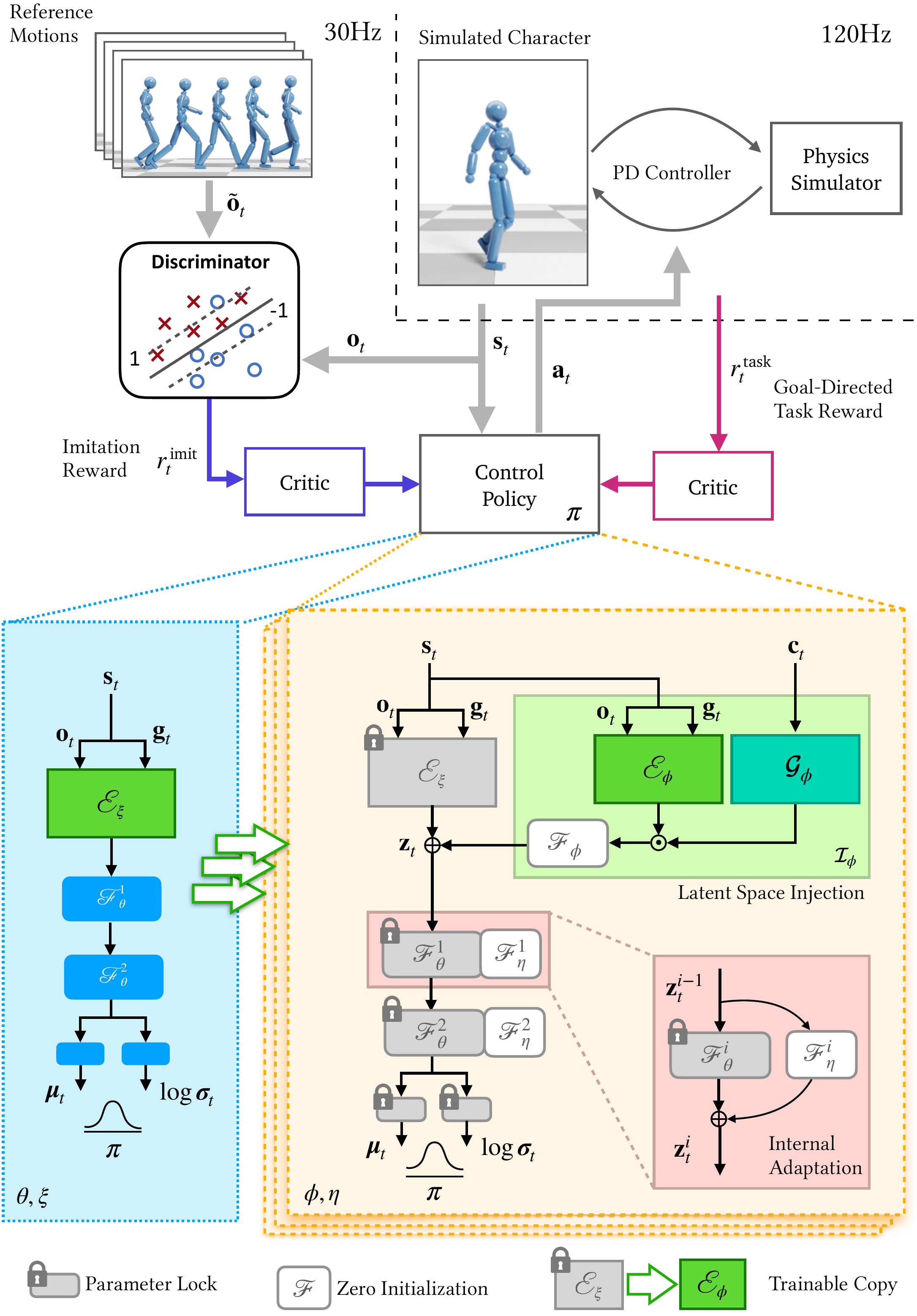}
    \caption{Overview of our approach for 
    adapting motor control policies for physics-based characters. Top: We model both pretraining and adapted tasks using 
    a multi-critic reinforcement learning framework that balances the training of imitation and goal-directed control objectives. After a policy is trained, we can quickly adapt it to a new task using AdaptNet. Bottom: AdaptNet starts with a copy of the pre-trained policy network and modifies it through %
     editing the latent space conditioned on the character's state and introducing optional %
     adaptation modules for further finetuning.
    }
    \label{fig:overview}
\end{figure}

\section{AdaptNet Framework}

An overview of the AdaptNet framework is shown in Figure~\ref{fig:overview}. The GAN-style control framework (top), described below, produces an original (pre-trained) policy (bottom, left) while  
AdaptNet is used to adapt that pre-trained control policy to a new task controller (bottom, right). 
Notably, the adaptation process could involve changes to the reward function (e.g., motion stylization) or the state and dynamics model %
(e.g., character morphology and terrain adaptation).
Components of the AdaptNet for policy adaptation are shown: a latent space injection component  
and an internal adaptation component. %
The latent space injection performs %
policy adaption by editing the latent space, which is conditioned on the pre-trained policy's state as well as any additional state information, for example, for new tasks.
This component is trained to cooperate with the pre-trained policy by generating offsets to the original latent space instead of trying to learn how to generate latent variables for new tasks from scratch during adaptation
This leads to an efficient state-action exploration that starts from the pre-trained policy, instead of complete random exploration. 
Internal adaptation further tunes the policy by adding a branch to each internal fully-connected layer in the policy network.
This allows for more flexibility, enabling AdaptNet to shift away from the pre-trained policy and generate refinement through control actions that the pre-trained policy may not reach easily.

In our implementation, both the pre-trained policy and the adaptation are produced using a multi-objective learning framework~\cite{composite} combining reinforcement learning with a GAN-like structure for effective policy learning that accounts for both motion imitation and goal-directed control (see Figure~\ref{fig:overview}, top). During runtime, AdaptNet can be activated flexibly and dynamically allowing us to control the level of adaptation of the original control policy. %
The control policy $\pi(\mathbf{a}\vert\mathbf{s}_t)$ is a 
neural network  
taking the agent state $\mathbf{s}_t$ as input and outputting a probability distribution from which a control %
$\mathbf{a}_t$ can be drawn from the action space $\mathcal{A}$. 
For physics-based character 
control tasks with dynamic goals, we consider $\mathbf{s}_t := \{\mathbf{o}_t, \mathbf{g}_t\}$, where $\mathbf{o}_t$ denotes the current state of the character, e.g., joint or body link positions and velocities, and $\mathbf{g}_t$ is an optional task-related goal state or an encoding variable that indicates 
desired motion parameters, such as target speed and direction, end-effector positions, motion style, etc.
The action vector $\mathbf{a}_t$ is %
the target posture fed to a PD servo through which the simulated character is controlled at a higher frequency. 
As shown in Figure~\ref{fig:overview}, $\mathbf{a}_t$ is expressed as a multivariate Gaussian distribution. 

Under the framework of reinforcement learning, 
our goal is to find the policy $\pi$ that maximizes the discounted cumulative reward: 
\begin{equation}
J = \mathbb{E}_{\tau \sim p(\tau\vert \pi)} \left[\sum_{t=0}\gamma^tr(\mathbf{s}_t,\mathbf{a}_t)\right], 
\end{equation}
where $p(\tau \vert \pi) = p(\mathbf{s}_0)\prod_{t=0}^{H-1} p(\mathbf{s}_{t+1}\vert \mathbf{s}_t, \mathbf{a}_t)\pi(\mathbf{a}_t \vert \mathbf{s}_t)$ is the state-action visitation distribution for the trajectory $\tau = \{s_t, a_t\}$ over a horizon of $H$ time steps,  
$\gamma \in [0,1]$ denotes the discount factor, 
and $r(\cdot)$ is the reward received at a given time step %
and $p(\cdot)$ is the state-transition probability of the underlying Markov decision process. 
In our domain, when the character faces a new task, $p(\cdot)$ and/or $r(\cdot)$ may change. 
AdaptNet seeks to efficiently modify $\pi$ and adapt it to the new task by 
editing the latent space and finetuning the policy.

\section{Policy Adaptation using Latent Space Injection}\label{sec:method_inject}

If we consider the first layer, or first several layers, in the policy network $\pi$ as an encoder to embed the state $\mathbf{s}_t$ into a latent space $\mathcal{Z}$, the control policy 
can be rewritten as
\begin{equation}
    \pi_\theta(\mathbf{a}_t \vert \mathcal{E}_\xi(\mathbf{s}_t)), 
\end{equation}
where $\mathcal{E}_\xi$ is the encoding layers with parameters $\xi$, $\theta$ are the parameters for the layers in the policy network that follow the encoder, and ($\theta, \xi$) denote the weights of $\pi$. 
In this formulation, 
the policy network $\pi_\theta$ decides the projection from the latent $\mathbf{z}_t = \mathcal{E}_\xi(\mathbf{s}_t)$ into the action space $\mathcal{A}$.
Assuming that %
$\pi_\theta$ 
is optimized by a typical on-policy policy gradient algorithm, the optimization objective with the introduction of the latent becomes 
\begin{equation}\label{eq:rl_loss}
    \max_{\theta,\xi} \mathbb{E}_t \left[A(\mathbf{s}_t, \mathbf{a}_t) \log\pi_\theta(\mathbf{a}_t \vert
    z_t; \xi)\right],
\end{equation}
where $A(\cdot)$ provides an advantage function estimation based on the received rewards $\{r_k\}_{k\geq t}$ during the interaction with the environment and represents how good an action sample $\mathbf{a}_t$ is given the conditional state $\mathbf{s}_t$. 

Given the generalization of neural networks, 
the latent space $\mathcal{Z}$ can be considered as a superset covering all the possible latent states,  
which could lie outside of the domain that $\pi_\theta$ can reach during its training.  

Based on this observation, when $\pi_\theta$ needs to be adapted to a new task, 
we propose to edit  $\mathbf{z}_t = \mathcal{E}_\xi(\mathbf{s}_t) \subset \mathcal{Z}$ instead of discarding the original encoder $\mathcal{E}_\xi$ and training a new one from scratch. The intuition is that for similar tasks, adjusting the current encoder provides %
better efficiency, allowing the desired control policy to be learned by a modified projection function from $\mathbf{s}_t$ to $\mathbf{z}_t$.

Our approach manipulates %
the full latent space projected from both the character state $\mathbf{o}_t$ and the goal state $\mathbf{g}_t$.
Specifically, as shown in Figure~\ref{fig:overview},
we perform latent space injection by introducing a new conditional encoder $\mathcal{I}_\phi$ with parameters $\phi$ after the first encoding layer, where the character state $\mathbf{o}_t$ and the goal state $\mathbf{g}_t$ are concatenated to generate %
$\mathcal{E}_\xi$. 
This latent space is modified via
\begin{equation}\label{eq:latent_after_injection}
\mathbf{z}_t = \mathcal{E}_\xi(\mathbf{s}_t) + \mathcal{I}_\phi(\mathbf{s}_t, \mathbf{c}_t), 
\end{equation}
where $c_t$ is an additional control input for the new task which could be optional.
The %
injector module $\mathcal{I}_\phi$ is %
\begin{equation}\label{eq:injection_component}
    \mathcal{I}_\phi(\mathbf{s}_t, \mathbf{c}_t) = \mathcal{F}_\phi(\textsc{Concat}(\mathcal{E}_\phi(\mathbf{s}_t), \mathcal{G}_\phi(\mathbf{c_t}))),
\end{equation}
where $\mathcal{G}_\phi$ is an optional module to process the additional control input $\mathbf{c}_t$, $\mathcal{E}_\phi$ is a state encoder that has exactly the same structure as the original encoder $\mathcal{E}_\xi$, and $\mathcal{F}_\phi$ is a final embedding module, which can be a fully-connected layer 
or a stack of multiple fully-connected layers. 

During retraining for adaptation, we perform policy optimization as in %
Eq.~\ref{eq:rl_loss}, but only optimize the new parameters $\phi$ while keeping the parameters $\theta$ and $\xi$ fixed:    
\begin{equation}\label{eq:injection_loss}
    \max_\phi \mathbb{E}_t \left[A(\mathbf{s}_t, \mathbf{a}_t) \log\pi_\theta(\mathbf{a}_t \vert \mathcal{E}_\xi(\mathbf{s}_t) + \mathcal{I}_\phi(\mathbf{s}_t, \mathbf{c}_t))\right].
\end{equation}
We begin with copying the original encoder parameters $\xi$ into the new encoder $\mathcal{E}_\phi$ and initializing the last fully-connected layer inside $\mathcal{F}_\phi$ with zero weight and bias.
In this way,
the new encoder $\mathcal{E}_\phi$ is optimized by finetuning a set of parameters that are already optimized for state feature extraction during pre-training.
The zero initialization of $\mathcal{F}_\phi$ lets the control policy give exactly the same action output as the original pre-trained one, i.e., $\pi_\theta(\mathbf{a}_t \vert \mathcal{E}_\xi(\mathbf{s}_t))$, at the beginning of re-training.
It guides the adaptation to start from the state-action trajectory generated by the original policy rather than from a completely random exploration.

We refer to Figure~\ref{fig:overview} for 
the default implementation of AdaptNet, where the latent space injection is performed right after the concatenation of $\mathbf{o}_t$ and $\mathbf{g}_t$.
We denote this latent space as $\mathcal{Z}^0$, and the following ones after each fully-connected layer but before the final action layer %
as $\mathcal{Z}^i$  where $i=1,2,\cdots$.
Empirically, we note that it is
more challenging to perform optimization 
when the injection occurs 
at a deeper layer in the policy network, leading typically to unstable training and low-fidelity controllers.  
An extreme case is to perform injection directly at the action space, which makes the whole system similar to directly finetuning the pre-trained policy network.
We refer to Section~\ref{sec:comparison} for related sensitivity analysis on introducing latent space injection at different %
network layers 
and 
for comparisons with directly finetuning a pre-trained policy network for new tasks.

During runtime, we can further introduce an extra scaling coefficient to the injection term in Eq.~\ref{eq:latent_after_injection}.
Since our approach does not change the original encoder $\mathcal{E}_\xi$ as well as the policy $\pi_\theta$, 
the scale coefficient allows us to turn the injection on and off, or
control the transition from the original policy to the fully adapted one. %
In such a way, we can perform %
motion style or behavior transitions (e.g., walk to skip) by interpolation in the latent space,
as we will show in Section~\ref{sec:style_transfer}.

\section{Internal Adaptation for Control Layers}\label{sec:method_adapt}

The latent space injection component of AdaptNet edits the latent space based on the input state and further allows us to introduce additional control input for new tasks.
However, the expressive ability of the action policy is still constrained by the pre-trained layers after the state encoder in the policy network, i.e., $\pi_\theta$. 
While utilizing the pre-trained $\pi_\theta$ for fast adaptation to new tasks, we introduce an internal adaptation component through which we can finetune $\pi_\theta$, 
overcoming the bias it introduces and allowing for more flexibility in the types of generated controls 
compared to the ones obtained from the original training domain. %
The goal of the %
finetuning is to find a \emph{small} increment $\Delta \mathbf{z}_t^i$ to the original latent $\mathbf{z}_t^i$ in each latent space $\mathcal{Z}^i,  i>1$, %
to help optimize the objective function in Eq.~\ref{eq:injection_loss} during adaptation training, but without changing the $\pi_\theta$ too much 
to avoid drifting too far away from the pre-trained policy and being stuck at overfitting during adaptation. 
To do so, we add a branch to each fully-connected layer between two latent spaces. 
As shown in the red block of Figure~\ref{fig:overview}, the corresponding latent is  generated as:
\begin{equation}\label{eq:adapt_component}
    \mathbf{z}_{t}^i = \mathcal{F}_{\theta}^i (\mathbf{z}_t^{i-1}) + \mathcal{F}_{\eta}^i(\mathbf{z}_t^{i-1}). 
\end{equation}
Here, $\mathcal{F}_{\theta}^i$ denotes the fully-connected layer between the latent space $\mathcal{Z}^{i-1}$ and $\mathcal{Z}^i$ in the policy network $\pi_\theta$, and $\mathcal{F}_{\eta}^i$ is the newly introduced adaptor that generates $\Delta \mathbf{z}_t^i$ and is modeled as a fully-connected layer in the added branch.
The parameter $\eta$ is defined as 
\begin{equation}
\label{eq:eta}
    \eta := \{\Delta\mathbf{W}_i, \Delta\mathbf{b}_i\}, 
\end{equation}
with $\Delta\mathbf{W}_i$ and $\Delta\mathbf{b}_i$ being the weight and bias parameters in $\mathcal{F}_{\eta}^i$ respectively.
Similarly to the embedding module $\mathcal{F}_\phi$ in the latent space injection component,
$\mathcal{F}_{\eta}^i$ is initialized as zero and will not influence the output of the policy network at the beginning of policy adaptation.
We lock $\theta$ in $\mathcal{F}_\theta^i$ during adaptation training and introduce the parameter $\eta$ into the optimization function in Eq.~\ref{eq:injection_loss}.

Our approach is different from directly finetuning $\pi_\theta$.
When directly finetuning $\pi_\theta$,
the gradient from $\mathbf{z}_t^i$ with respect to $\mathbf{z}_t^{i-1}$ is decided by the weight $\mathbf{W}_i$ in the layer $\mathcal{F}_\theta^i$, which may be highly biased and have relatively large or very small values given it was fully trained.
Therefore, finetuning $\pi_\theta$ directly for new tasks may lead to unstable training compared to only finetuning the newly introduced parameter set $\eta$ which is initialized with zero. %
Furthermore, we can easily apply regularization on $\Delta\mathbf{W}_i$ and $\Delta\mathbf{b}_i$ to prevent aggressive finetuning regardless of the value of the parameters $\mathbf{W}_i$ and $\mathbf{b}_i$ in the pre-trained layer $\mathcal{F}_\theta^i$.
This will limit the possible change that the internal adaptation can bring about
in order to prevent overfitting.
We can also introduce an extra scaling weight to control the adaptation level during runtime, as discussed in Section~\ref{sec:method_inject}.

Our proposed internal adaptation component is similar to the approach of low-rank adaptation (LoRA) proposed by~\citet{hu2021lora}.
The major difference %
is that instead of directly employing a fully-connected layer,
LoRA decomposes the weight matrix $\Delta \mathbf{W}_i$ into two low-rank matrices, i.e.,
    $\Delta \mathbf{W}_i = \mathbf{B}_i \mathbf{A}_i$, 
where, %
$\mathbf{B}_i$ is a $\vert \mathcal{Z}^{i-1} \vert$-by-$r$ matrix, 
$\mathbf{A}_i$ is a $r$-by-$\vert \mathcal{Z}^i \vert$ matrix,
and $r \ll \min(\vert \mathcal{Z}^{i-1} \vert, \vert \mathcal{Z}^i \vert)$. 
In contrast, our approach can be considered a full-rank adaptation. 
LoRA has been demonstrated as an effective way to fine tune large language and image generation models, reducing the number of parameters that need to be optimized during model adaptation.
However, as shown in Section~\ref{sec:lora},
we found that %
LoRA does not work well for physics-based character control tasks.
A possible reason is that the related policy networks are markedly
smaller compared to large language and image generation models that may have more than 12K dimensions. 
The latent spaces of our policy network have a typical size of 512 or 1024 dimensions and may 
not exhibit the
lower intrinsic ranks that %
larger models do~\cite{aghajanyan2020intrinsic,pope2021the,li2018measuring}.

\section{Policy Training}\label{sec:method_training}

\begin{algorithm}[t]
\emph{Obtain the policy $\pi_{\theta}$ and the state encoder $\mathcal{E}_\xi$ by performing training to optimize  Eq.~\ref{eq:loss_pretraining} in a general or default environment setting.} %

\nl Build up the latent space injection component $\mathcal{I}_\phi$ based on Eq.~\ref{eq:injection_component} and the internal adaptation component $\{\mathcal{F}_\eta^i\}$ based on the Eq.~\ref{eq:adapt_component}.

\nl Lock the parameters $\theta$ and $\xi$.

\nl Initialize $\mathcal{E}_\phi$ using the pre-trained parameter $\xi$.

\nl Initialize the last layer inside $\mathcal{F}_\phi$ and each $\mathcal{F}_\eta^i$ using zero weight and bias.

\nl Adapt the policy for a new task by only optimizing the parameters $\phi$ and $\eta$ using Eq.~\ref{eq:loss_adapt}.
\caption{Policy Adaptation using AdaptNet}
\label{alg:alg}
\end{algorithm}

We use the multi-objective learning framework  for physics-based character control proposed by~\citet{composite} to perform both
the original (pre-)training and adaptation training. 
The framework leverages a 
multi-critic structure where the 
objectives of motion imitation and goal-directed control are considered independent tasks during policy updating.  
In Figure~\ref{fig:overview}, for example,
the imitation objective is associated with a critic network labeled in blue, and the goal-directed objective is associated with a critic in magenta.
The advantage (cf. Eqs.~\ref{eq:rl_loss},~\ref{eq:injection_loss}) with respect to each objective is estimated only by its associated reward and critic network.
To ensure that the policy can be updated in a balanced way taking into account both the imitation and goal-directed control objectives, 
all estimated advantages are standardized independently before policy updating. 

During pre-training, we seek to find a basic motor control %
policy 
$\pi_\theta(\mathbf{a}_t \vert \mathcal{E}_\xi(\mathbf{s}_t))$, 
which we can later adapt to new tasks. 
In this work, we focus on locomotion tasks, and thus $\pi_\theta$ involves two objectives: a motion imitation objective given a batch of reference motions of walking and running, and a goal-directed objective involving a given target direction and speed. %
Using the multi-objective learning framework, the optimization objective function during pretraining shown in Eq.~\ref{eq:rl_loss} can be written as 
\begin{equation}\label{eq:loss_pretraining}\begin{split}
    \max_{\theta, \xi} \mathbb{E}_t \Bigl[
    \Bigl(\sum_k \omega_k \bar{A}_t^k\Bigr)
    \log\pi_\theta\bigl(\mathbf{a}_t \vert \mathcal{E}_\xi(\mathbf{s}_t)\bigr) \Bigr],
\end{split}\end{equation}
where $\bar{A}_t^k$ is the standardization of the estimated advantage associated with the objective $k$ and $\omega_k$ satisfies $\sum_k \omega_k = 1$ providing additional control to adjust the policy updating in a preferred manner when conflicts between multiple objectives occur. %

We employ a GAN-like structure~\cite{ho2016generative,merel2017learning} that relies on an ensemble of discriminators~\cite{iccgan} to evaluate the imitation performance and generate the corresponding reward signals for advantage estimation and policy updating.
In particular, 
we take an ensemble of $N$ discriminators and use a hinge loss~\cite{lim2017geometric} with policy gradient~\cite{gulrajani2017improved} for discriminator training, resulting in the following loss function:  
\begin{equation}\label{eq:dis_loss}\begin{split}
    \min \frac{1}{N}\sum_{n=1}^{N}\Bigl( \mathrm{E}_t\left[\max(0, 1+D_n(\mathbf{o}_t))\right] + \mathrm{E}_t\left[\max(0, 1-D_n(\mathbf{\tilde{o}}_t))\right] \\
    + \lambda^{\text{GP}} \mathrm{E}_t\left[(\| \nabla_{\mathbf{\hat{o}}_t} D_n(\mathbf{\hat{o}}_t) \|_2 - 1)^2\right]\Bigr). \qquad
\end{split}\end{equation}
Here, $D_n$ denotes a discriminator network, $\mathbf{\hat{o}}_t = \alpha \mathbf{o}_t + (1-\alpha) \mathbf{\tilde{o}}_t$ with $\alpha \sim \textsc{Uniform}(0, 1)$
and $\lambda^{\text{GP}}$ is gradient penalty coefficient. 
The reward function to evaluate the imitation performance is defined as
\begin{equation}\label{eq:dis_rew}
    r^\text{imit}(\mathbf{s}_t, \mathbf{a}_t, \mathbf{s}_{t+1}) = \frac{1}{N} \sum_{n=1}^N \textsc{Clip}\left(D_n(\mathbf{o}_t), -1, 1\right).
\end{equation}

\noindent The reward for the goal-related task is computed heuristically. We refer to the appendix %
for the representation of the goal state $\mathbf{g}_t$ and the definition of the goal-related task reward.

After obtaining $\pi_\theta$ and $\mathcal{E}_\xi$ %
in pre-training,
we introduce the proposed AdaptNet to perform policy adaptation for new tasks that are relative to but have different reward definitions and/or environment settings from the one in the pre-training phase. 
Before the adaptation training starts, we lock the parameters $\theta$ and $\xi$. We then initialize $\mathcal{E}_\phi$ inside the latent space injection component $\mathcal{I}_\phi$  using the %
weights $\xi$, and initialize with zero weight and bias 
the last layer of $\mathcal{F}_\phi$ inside $\mathcal{I}_\phi$ along with each fully-rank adaptor $\mathcal{F}_\eta^i, i>0$. %
To stabilize the training, 
besides applying a common weight decay to the parameter set $\eta$ (Eq.~\ref{eq:adapt_component}) via L2 regularization,
we introduce an additional regularization on the latent injection generated by $\mathcal{I}_\phi$.
The adaptation training is still performed under the aforementioned multi-objective learning framework in the same way as the pre-training phase.
The optimization objective for policy adaptation is
\begin{equation}\label{eq:loss_adapt}\begin{split}
    \max_{\phi, \eta} \mathbb{E}_t \Bigl[
    \Bigl(\sum_k \omega_k \bar{A}_t^k\Bigr)
    \log\pi_\theta\bigl(\mathbf{a}_t \vert \mathcal{E}_\xi(\mathbf{s}_t) + \mathcal{I}_\phi(\mathbf{s}_t, \mathbf{c}_t); \eta\bigr) \\
    - \beta \| \mathcal{I}_\phi(\mathbf{s}_t, \mathbf{c}_t)\|_2 - \kappa \| \eta \|_2 \Bigr], \qquad
\end{split}\end{equation}
where $\beta$ and $\kappa$ are regularization coefficients.
In Section~\ref{sec:latent}, we give a detailed analysis of the regularization on the latent space injection. 

We refer to Algorithm~\ref{alg:alg} for the outline of the whole training process. %
Adaptation with the proposed AdaptNet can be done very quickly within 10-30 minutes for simple control tasks and up to 4 hours for %
challenging terrain adaptation tasks with  new control input processed by an additional convolutional neural network $\mathcal{G}_\phi$, as defined in Eq.~\ref{eq:injection_component}.

\section{Experimental Setup}\label{sec:setup}
\begin{figure}[t]
\centering
    \begin{subfigure}[t]{.3\linewidth}
    \includegraphics[width=\linewidth]{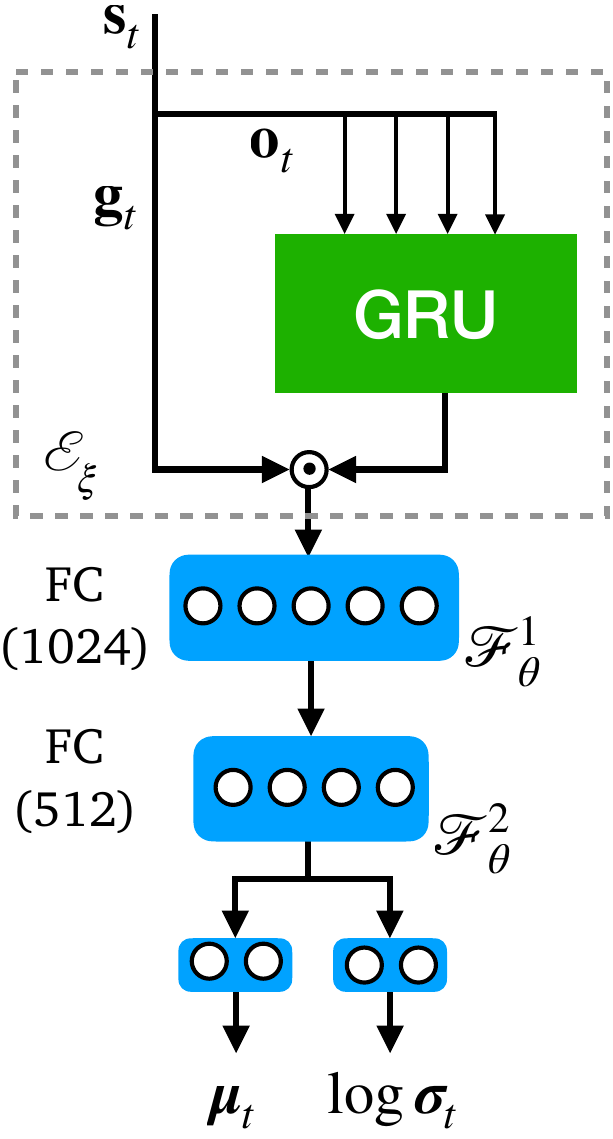}
    \caption{Policy Network}
    \end{subfigure}\hfill
    \begin{subfigure}[t]{.36\linewidth}
    \hfill\includegraphics[width=.97\linewidth]{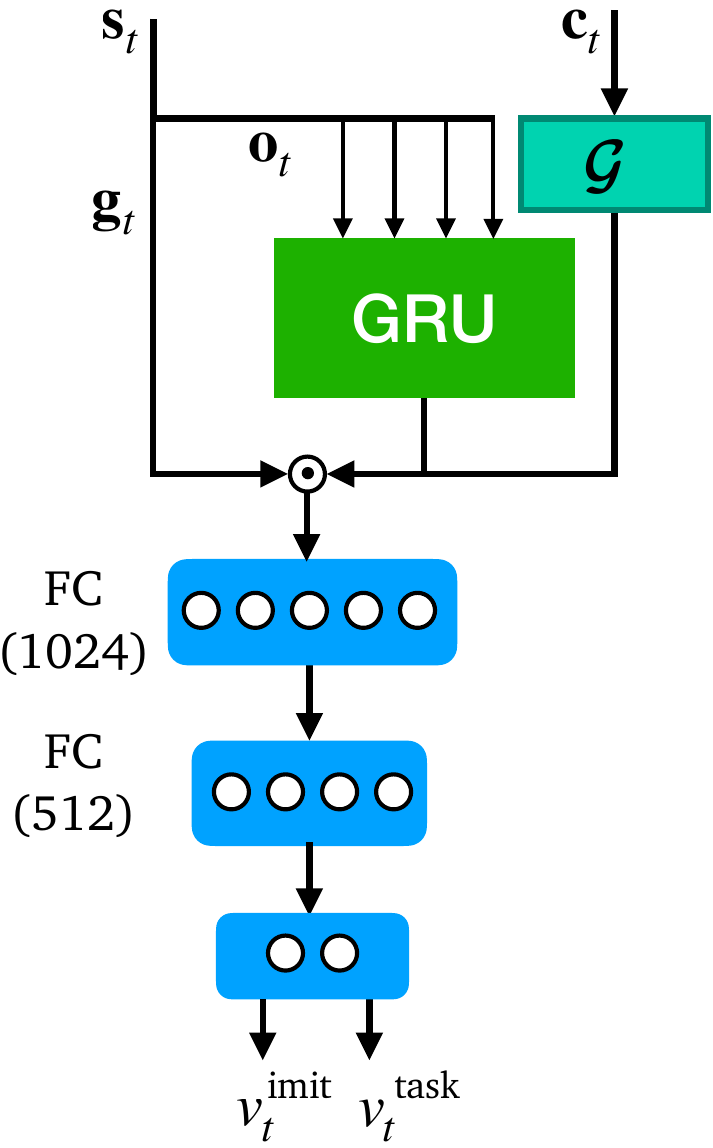}
    \caption{$2$-Head Value Network}
    \end{subfigure}\hfill
    \begin{subfigure}[t]{.3\linewidth}
    \centering
    \includegraphics[width=.75\linewidth]{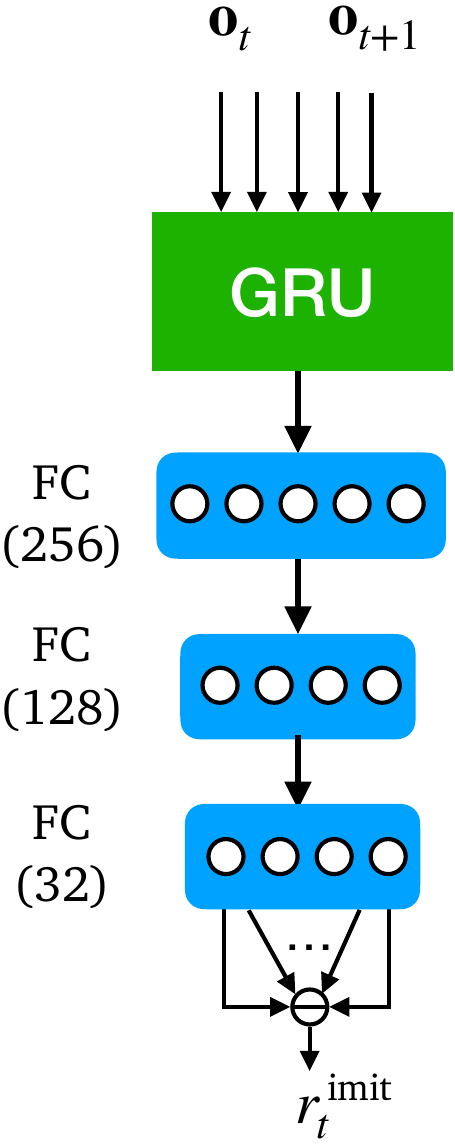}
    \caption{Discriminator}
    \end{subfigure}
    \caption{Network structures. Here, $\odot$ denotes the concatenation operator and $\ominus$ denotes the average operator. The state encoder $\mathcal{E}_\xi$ is shown in the dashed block. 
    An optional control input encoding module $\mathcal{G}$ is included if the additional control input $\mathbf{c}_t$ is provided during adaptation training.
    }
    \label{fig:network}
\end{figure}

Our experiments were run using %
IsaacGym~\cite{makoviychuk2021isaac} with 512 environments running in parallel during training. 
The simulated character has 15 body links and 28 degrees of freedom,
where the elbow and knee joints are implemented as 1-dimensional revolute joints, and the hands are fused with the forearms and uncontrollable.
All simulations run at 120Hz with a normal PD controller employed as the low-level 
actuator
to directly manipulate the simulated character, 
while the control policy runs at 30 Hz, as shown in Figure~\ref{fig:overview}.

We run policy optimization using PPO~\cite{schulman2017proximal} and update policy parameters using the Adam optimizer~\cite{kingma2014adam}.
To encode the character's state, %
we take the position, orientation, and velocities of all the body links related to the pelvis (root link) in the last four frames as the state representation $\mathbf{o}_t$ and employ a gated recurrent unit (GRU)~\cite{chung2014empirical} with a 256-dimension hidden state to process this temporal state.
For discriminator training, we take the character's pose at five consecutive frames as the representation of $\{\mathbf{o}_t, \mathbf{o}_{t+1}\}$ to evaluate the policy's imitation performance during the transition from timestep $t$ to $t+1$.
We employ an ensemble of 32 discriminators and model it by a multi-head network, as shown in Figure~\ref{fig:network}.
The critic network has a similar structure to the policy network, but with a 2-dimensional output for the value estimations to the imitation objective and goal-directed objective %
respectively. 
We refer to the appendix %
for the hyperparameters used for policy training and the representation of the goal state $\mathbf{g}_t$ in the locomotion task.

Rewards for both task and imitation are employed during policy adaptation.  To avoid bias from the pre-trained policy, we discard the discriminators for imitation from the original policy and new discriminators are trained from scratch. 
Intuitively, in tasks such as motion style transfer %
the original discriminator will not work well for the new given reference style and thus a new one is needed. Even for other adaptation tasks, we found utilizing old discriminators to be problematic, as the optimal action in the new task can dramatically change from the original in the context of  how it employs the reference motion.  
Empirically, when we experimented with reusing the old discriminators, we found they introduce too much bias towards the old task. Finally, with training new discriminators for a new task, we also perform value estimation by re-training a new critic from scratch.

All our tests were run on machines with a V100 or A100 GPU.
To achieve a good locomotion policy based on which we perform further adaptation,
the pre-training took around 26 hours and consumed  $~4\times10^8$ training samples. %
The reference motions are around 300 seconds long including normal walking and running motions with turning poses and various speeds (cf. Table~\ref{tab:motions}, top). 
All the reference motions used during pre-training and adaptation training  
are recorded at 30~Hz and extracted from the publicly available dataset LAFAN1~\cite{harvey2020robust}.

\section{Applications of AdaptNet }\label{sec:results}
In this section, we apply the AdaptNet technique to demonstrate the success and efficiency of learning new physics-based controllers through adaptation. 
Our experiments use two pre-trained locomotion policies (walking and running) that account for two objectives: motion imitation based on a batch of walking or running reference motions, respectively, and a goal objective as defined by a target direction of motion and speed. 
We adapt the pre-trained policies to a range of new tasks, highlighting 
applications of AdaptNet to 
style transfer,  
character morphology changes and 
adaptation to different terrains. 
Figure~\ref{fig:teaser} shows snapshots from different outcomes. Please refer to the supplementary video for related animation results.

\subsection{Motion Style Transfer and Interpolation}\label{sec:style_transfer}

\begin{table}[t]
    \centering\small 
    \caption{Reference motions for policy pre-training (top) and stylized motion learning (bottom).}
    \vspace{-8pt}
    \begin{tabular}{r|c|l}
        \toprule
        \textbf{Motion} & \textbf{Length} & \textbf{Description} \\
        \midrule
        Walk & 334.07~s & {\footnotesize normal walking motions for pre-training}\\
        Run & 282.87~s & {\footnotesize normal running motions for pre-training}\\
        \midrule
        Swaggering Walk & 1.07~s & {\footnotesize exaggerated walking with one arm akimbo} \\
        Goose Step & 2.20~s & {\footnotesize goose step with arms akimbo} \\
        Stomp Walk & 1.23~s & {\footnotesize walking while stomping on the ground} \\
        Kicking Walk & 2.03~s & {\footnotesize walking with leg kicking} \\ %
        Stoop & 0.93~s & {\footnotesize slow walking with body bent over} \\
        Jaunty Skip & 1.60~s & {\footnotesize skipping in a spirited manner} \\
        Sashay Walk & 1.07~s & {\footnotesize walking in a slightly exaggerated manner} \\
        Limp & 1.90~s & {\footnotesize slow walking with right leg hurt} \\
        Pace & 1.70~s & {\footnotesize slow walking with arms akimbo} \\
        Penguin Walk & 0.77~s & {\footnotesize moving with very small and steps} \\ %
        Strutting Walk & 1.40~s & {\footnotesize walking with shoulder moving aggressively } \\ %
        Joyful Walk & 1.20~s & {\footnotesize strut walking rhythmically} \\
        \bottomrule
    \end{tabular}
    \label{tab:motions}
\end{table}

\begin{figure*}[t]
    \centering
    \includegraphics[width=.202\linewidth]{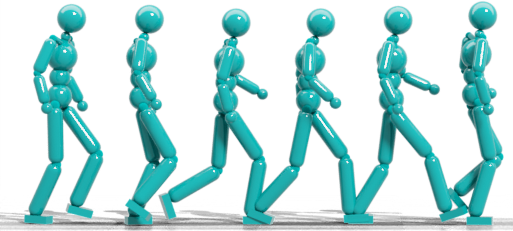}\hfill\includegraphics[width=.65\linewidth]{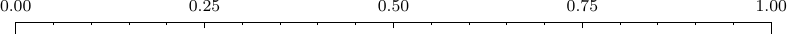}\qquad\qquad\qquad\qquad\qquad
    
    \includegraphics[width=\linewidth]{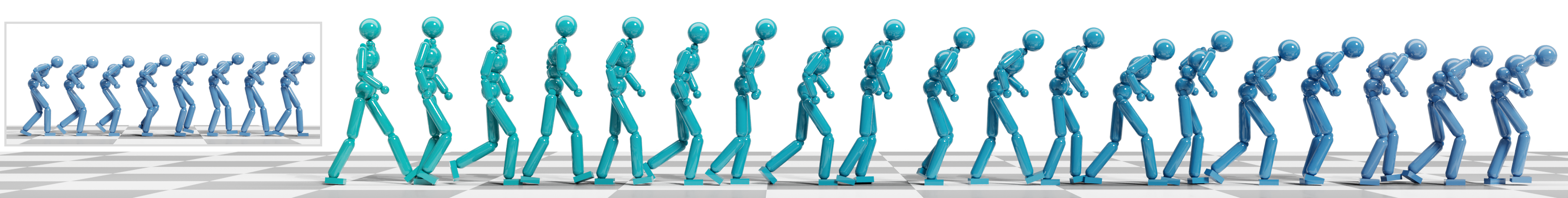}\\\vspace{0.2cm}
    \includegraphics[width=\linewidth]{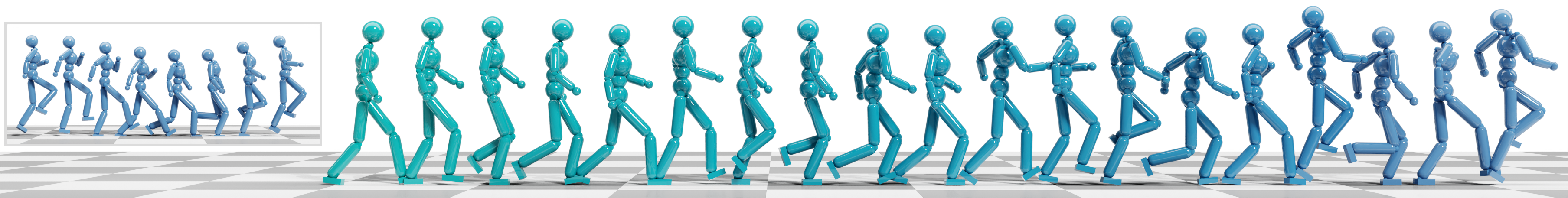}\\\vspace{0.2cm}
    
    \hfill\includegraphics[width=.65\linewidth]{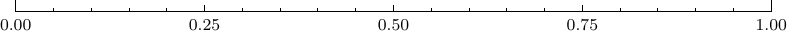}\qquad\qquad\qquad\qquad\qquad
    \caption{Motion interpolation between walking (pre-trained policy) shown at the top-left corner and different stylized motions by controlling the adaptation level of the associated AdaptNet model (cf. Eq.~\ref{eq:motion_interp}). 
    Snapshots on the left show the learned stylized motions of \textit{Stoop} walking and \textit{Jaunty Skip}.
    When $\alpha=0$, %
    the character is controlled only by the original walking policy. 
    When $\alpha=1$, 
    the character is controlled with a full injection of AdaptNet.
    }
    \label{fig:style_interp}
\end{figure*}

\begin{figure}[t]
    \centering
    \includegraphics[width=\linewidth]{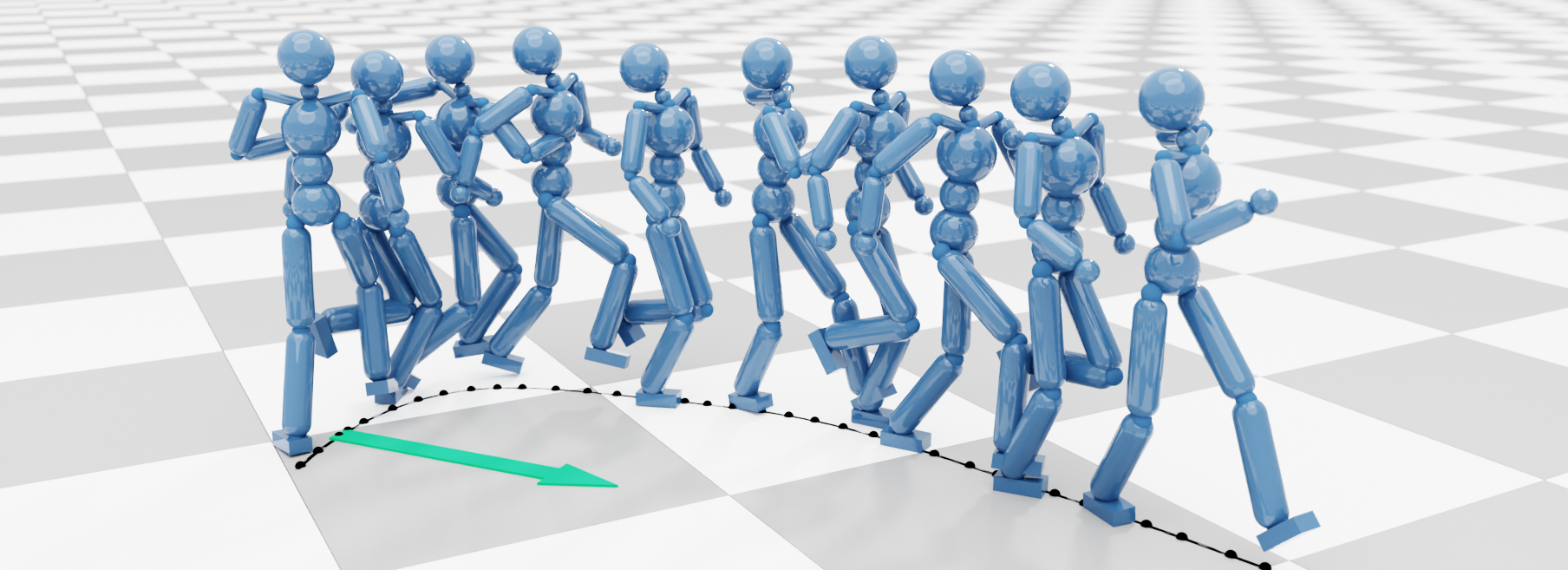}%
    \caption{Example of motion style transfer learning %
    with goal-steering navigation using AdaptNet.
    Green arrow indicates the dynamically generated target directions for locomotion control. 
}
    \label{fig:style_trans_nav}
\end{figure}

We consider a variety of 
motion style transfer tasks where a pre-trained walking locomotion policy is adapted to a particular style. 
Note, this is not a simple motion imitation task, since 
all the style reference motions %
are very short (see Table~\ref{tab:motions}, bottom),  containing only one or two gait cycles.  
It is therefore impossible to train an equivalent locomotion policy that supports goal-directed steering using the target reference motion. 
Instead, the nature of this test is few-shot learning, 
where AdaptNet is expected to effectively learn how to perform locomotion in the style provided by the small duration of the style
example in the new reference, 
while relying on the pre-trained policy to perform turning and goal-directed steering. 
Figure~\ref{fig:style_trans_nav} depicts related qualitative results. 
AdaptNet can effectively learn how to do goal-directed turning in the provided style.  
Further, adaptation training can be done very quickly, within 10-30 minutes, 
in contrast to 
the original that we obtained during pre-training took about one day for training.
We refer to the supplementary video for  animation results, and Section~\ref{sec:comparison} for comparing  AdaptNet to learning stylized locomotion from scratch.  %

As discussed in Sections~\ref{sec:method_inject} and~\ref{sec:method_adapt},
we can perform motion interpolation in the latent space by introducing a scale variable to control the adaptation level.
This process can be described by modifying Eqs.~\ref{eq:latent_after_injection} and~\ref{eq:adapt_component} as
\begin{equation}\label{eq:motion_interp}\begin{split}
     \mathbf{z}_t^{0} & = \mathcal{E}_\xi(\mathbf{s}_t) + \alpha \mathcal{I}_\phi(\mathbf{s}_t, \mathbf{c}_t), \\
     \mathbf{z}_t^i & = \mathcal{F}_{\theta}^i (\mathbf{z}_t^{i-1}) + \alpha \mathcal{F}_{\eta}^i(\mathbf{z}_t^{i-1}), 
\end{split}\end{equation}
where $\alpha \in [0, 1]$ is the introduced scale variable.
In Figure~\ref{fig:style_interp}, we show  interpolation results. 
As shown in the figure, 
we can achieve motions with different style intensity, which can transition between the base walking motion and the stylized ones in a smooth manner.

\begin{figure*}[t]
   \centering\vspace{10pt}
   \includegraphics[width=\linewidth]{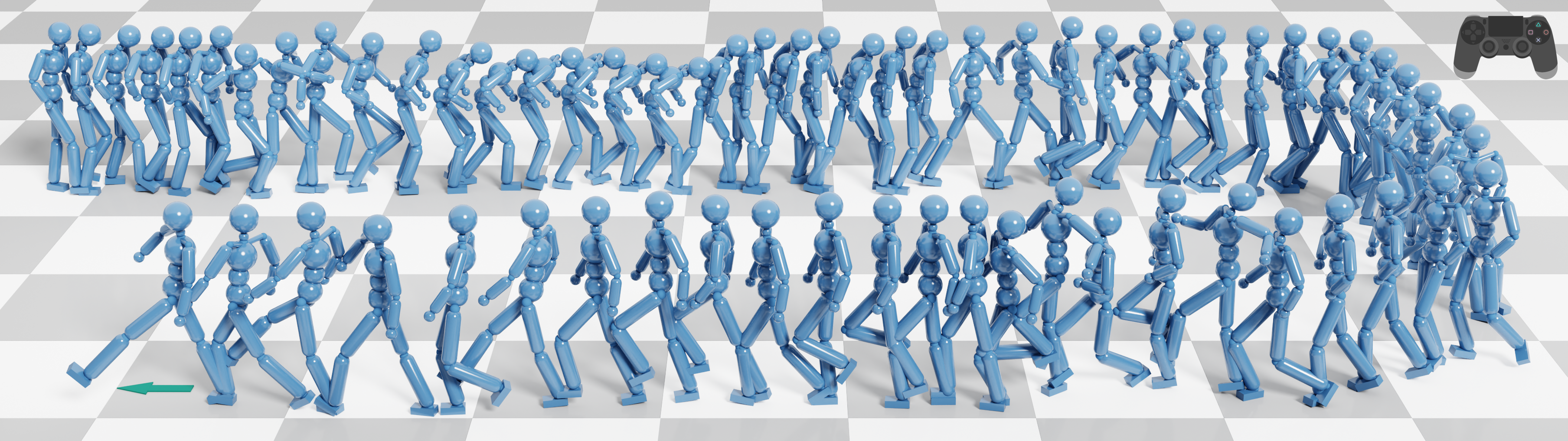}
   \caption{Motion interpolation in the latent space by activating and switching between multiple AdaptNet models to let the character perform style transition interactively during goal-steering navigation. %
   }
   \label{fig:style_interp_interactive}
\end{figure*}

We can further extend Eq.~\ref{eq:motion_interp} to perform interpolation between any two AdaptNet models via%
\begin{equation}\begin{split}
     \mathbf{z}_t^{0} & = \mathcal{E}_\xi(\mathbf{s}_t) + \alpha \mathcal{I}_{\phi^\prime}(\mathbf{s}_t, \mathbf{c}_t) + (1-\alpha) \mathcal{I}_{\phi^{\prime\prime}}(\mathbf{s}_t, \mathbf{c}_t), \\
     \mathbf{z}_t^i & = \mathcal{F}_{\theta}^i (\mathbf{z}_t^{i-1}) + \alpha \mathcal{F}_{\eta^\prime}^i(\mathbf{z}_t^{i-1}) + (1-\alpha)\mathcal{F}_{\eta^{\prime\prime}}^i(\mathbf{z}_t^{i-1}),
\end{split}\end{equation}
where the parameters $\phi^\prime$ and $\eta^\prime$ are from one AdaptNet model and $\phi^{\prime\prime}$ and $\eta^{\prime\prime}$ are from the other one.
Such an interpolation scheme can be regarded as applying two independently trained AdaptNet models simultaneously on the same, pre-trained policy, with an example shown in 
Figure~\ref{fig:style_interp_interactive}.

The above interpolation results demonstrate 
that during adaptation training, AdaptNet can effectively learn structured information about the latent space with respect to the desired motion styles.
We refer to Section~\ref{sec:latent} for more details on controlling the latent space and related visualizations,  
along  
with an analysis of the training difficulty (time consumption) when learning different styles.

\subsection{Morphological Adaptation}
We consider two kinds of morphological changes:
body shape %
and joint lock.
Due to physical constraints,
morphological changes in the character model will cause
the same action $\mathbf{a}_t$ to lead to different resulting states compared to the ones observed in the pre-training phase.
Without adaptation, the pre-trained policy does not perform well if it's even able to keep the character balanced, %
especially when the lower body is modified.

\begin{figure}[t]
    \centering
    \includegraphics[width=\linewidth]{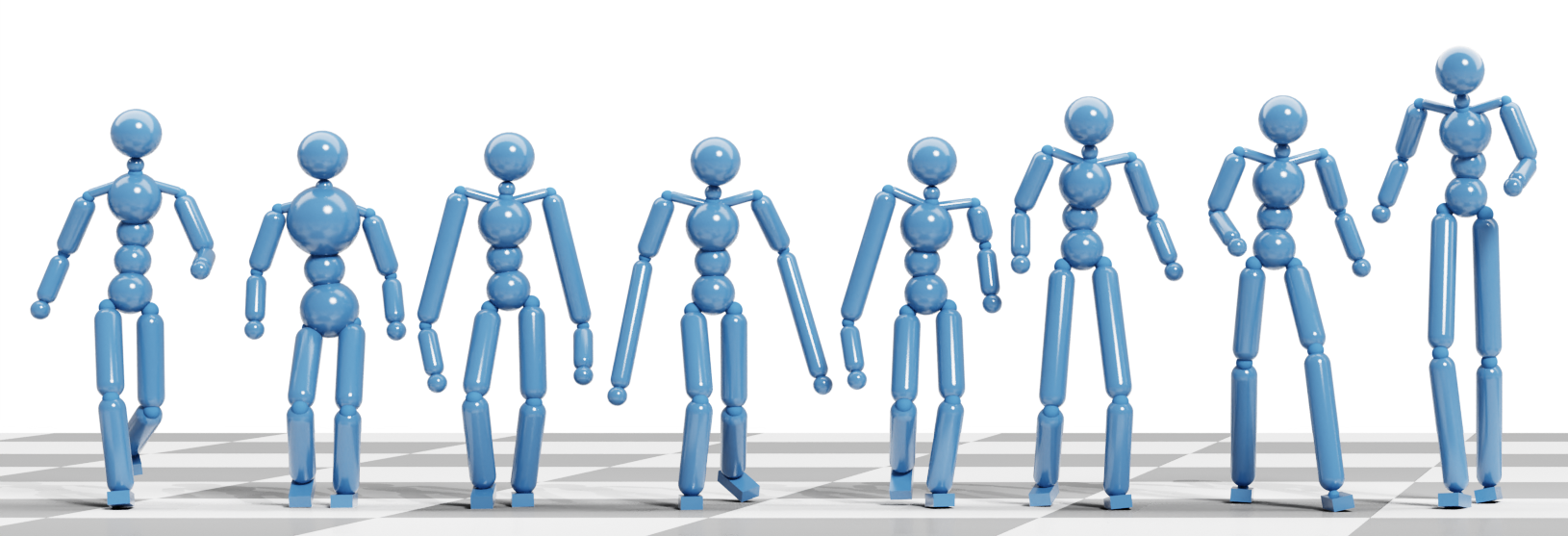}
    \caption{Character models with body shape variants. From left to right: \textit{LongBody}, \textit{BigBody}, \textit{LongUpperArms}, \textit{LongLowerArms}, \textit{AsymmetricUpperArms}, \textit{LongThighs}, \textit{LongShins}, and \textit{SuperLongLegs}.}
    \label{fig:bodyshape_showcase}
\end{figure}
\begin{figure}[t]
    \centering
    \includegraphics[width=\linewidth]{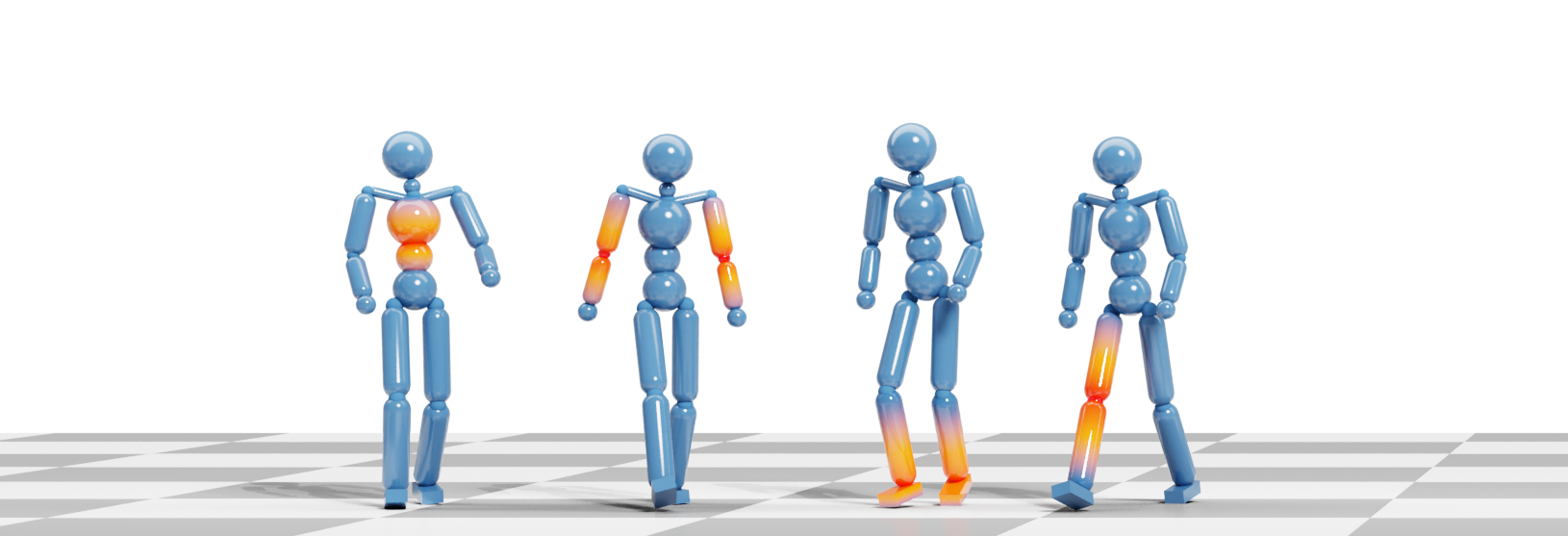}
    \caption{Character models with joints being locked. 
    From left to right, the locked joints are abdomen, elbows, ankles, and right knee respectively (shown in red).
    Corresponding body parts between a locked joint are highlighted in orange. %
    }
    \label{fig:jointlock_showcase}
\end{figure}
\begin{figure*}[t]

    \centering
    \includegraphics[width=\linewidth]{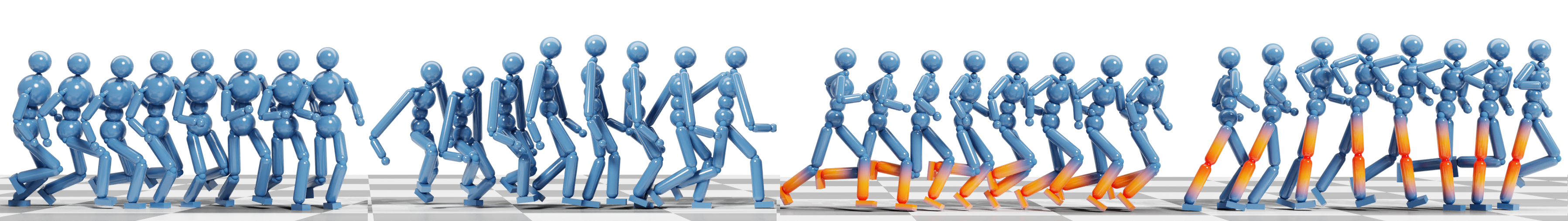}
    \caption{Adapting the locomotion policy of running to characters with different body shapes and locked joints. 
    }
    \label{fig:morph}
\end{figure*}

We tested eight body-shape variants of the original character model, as shown in Figure~\ref{fig:bodyshape_showcase}.
In the \textit{LongBody} variant, we extend the abdomen length by 50\%, 
while the \textit{BigBody} variant increased the torso size by 50\%.
The latter leads to an increase in the torso mass of over 300\%.
In \textit{LongUpperArms} and \textit{LongLowerArms} variants, the length of the upper and lower arms are extended by 25\% respectively,
while in \textit{AsymmetricUpperArms}, we increase the length of the right upper arm 
but decrease the length of the left upper arm.
In the \textit{LongThighs} and \textit{LongShins} variants, the length of the upper and lower legs are extended by 50\% respectively,
the latter akin to a human walking on stilts.
In the model of \textit{SuperLongLegs}, both the thighs and shins are extended resulting in a character that is over 2~m tall.%

We also experimented with different configurations, as shown in Figure~\ref{fig:jointlock_showcase}, 
where some of the joints (in orange) are `locked'.
The locked joints are removed from the character model such that the linked body parts are %
fused together.
This reduces the number of dimensions of the action space. 
To make the pre-trained policy compatible with the new action space, %
we simply prune the weight and bias matrices of the last layers in the policy network
and remove the output neurons corresponding to the locked joints.

Even though the pre-trained policy would not completely lose control of the character when  %
the torso or arms are modified,
the character %
still loses balance quite often.
As more challenging examples,
the morphological changes in the lower body parts and joints
leave the pre-trained policy unable to control the character without falling.

For example, when the knee joint is locked, %
the policy needs to adjust the output of the hip and ankle in order to compensate for the `disability' of the knee.
This requirement leaves the pre-trained policy incapable of suitably controlling the modified character model.%

During adaptation, we did not do any retargeting to generate new reference motions for AdaptNet to learn.
Instead, %
we simply modify the character's model while relying on the reference motions used to pre-train the original policy, retargeted to the character model without any morphological changes. %
We found it takes 15-30 minutes to finish the adaptation training depending on the difficulty of the morphology change task.
The character controlled by the AdaptNet policy can maintain its balance and walk or run without falling down. 
An interesting observation is that 
in order to match the provided height of the root link (pelvis) in the reference motions, the
AdaptNet policy will control the character to walk or run in a crouch with the body at a relatively low position compared to the leg length.
We show some representative %
results in Figure~\ref{fig:morph},
and refer to the supplementary video for animations. %

\subsection{Terrain Adaptation}
\label{sec:domainadapt}
\begin{figure*}[t]
    \centering
    \includegraphics[width=0.66\linewidth]{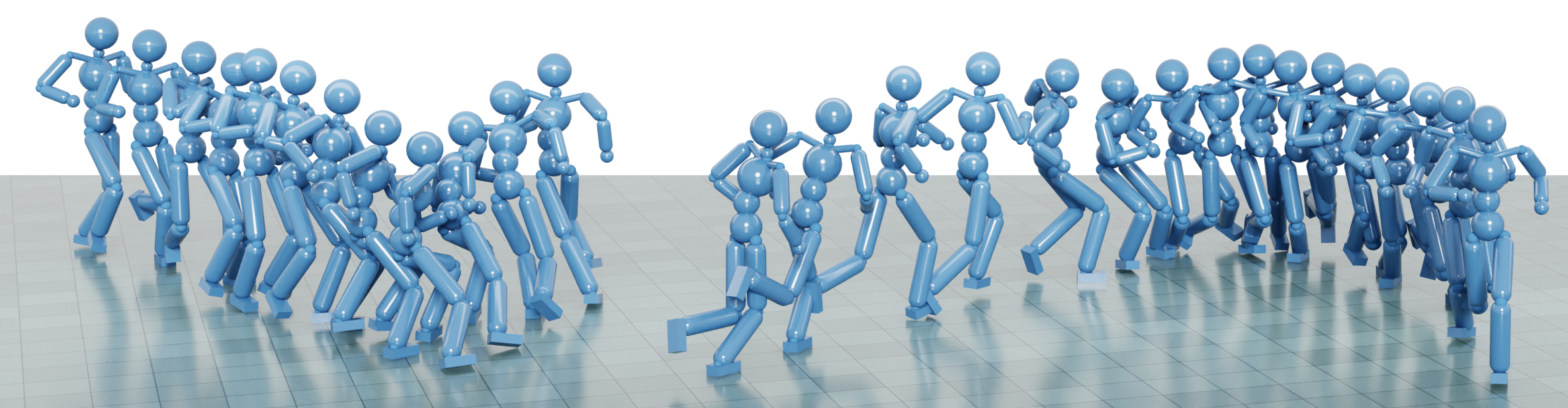}\hfill
    \includegraphics[width=0.33\linewidth]{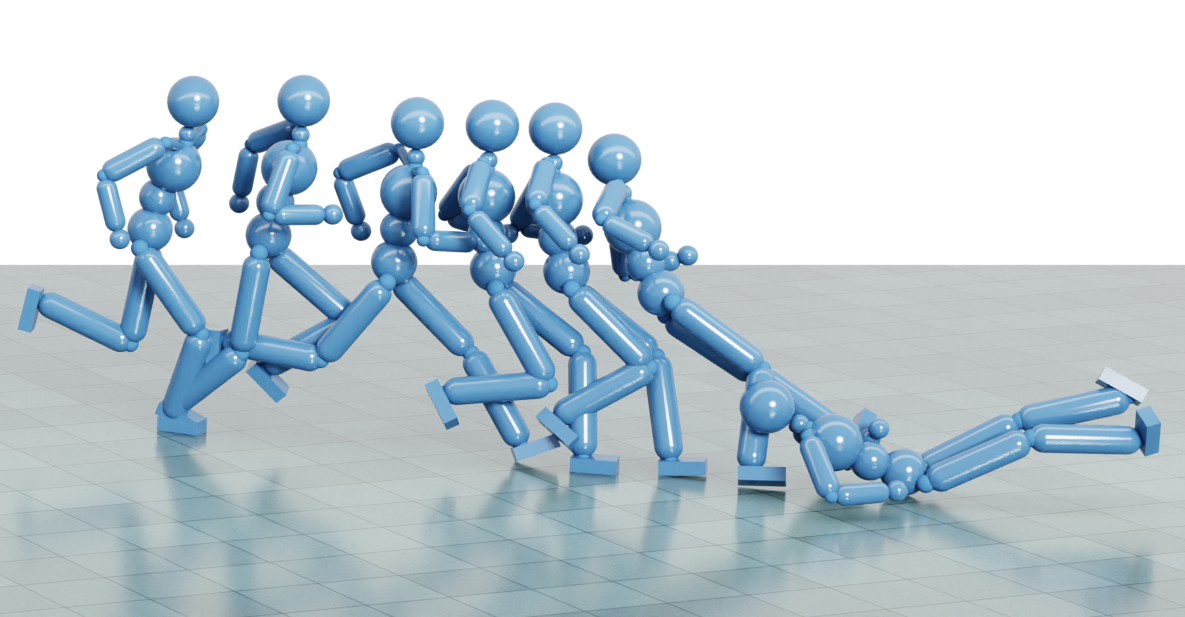}
    \caption{Comparison of characters controlled with and without AdaptNet running on an ice floor with very low friction. Left: character controlled with AdaptNet slides and skids on the ice ground while running. Right: character without AdaptNet slips down. 
    }
    \label{fig:friction}
\end{figure*}

Next we discuss policy adaptation for character locomotion on low friction and rough terrains as well as obstacle-filled scenes that require extra control input.

\subsubsection{Friction Adaptation}
To simulate an icy surface,
we significantly reduce the ground friction.
In particular, we decrease the friction coefficient  
from 1 to 0.15 for walking and to 0.35 for running. 
Figure~\ref{fig:friction} compares results obtained for the running policy with and without using AdaptNet. 
Note, AdaptNet can effectively control the character to change its moving direction by sliding on its feet, 
as shown in the left example of the figure. 
In addition, using AdaptNet, 
the character lowers its center of mass and takes quick steps to maintain its balance. 
In contrast, 
with the original %
policy, the character cannot run on the icy ground without falling down. 
For walking, the AdaptNet controller is more cautious with the character preferring to stop and change its direction in place. 
Without using AdaptNet, the character tends to turn around with a bigger radius, but not slow down. 
This demonstrates the ability of AdaptNet to change the behavior provided by the original policy to make it better suited to new environmental settings.

\subsubsection{Terrain Adaptation with Additional Control Input}
\begin{figure}[t]
    \centering
    \includegraphics[width=\linewidth]{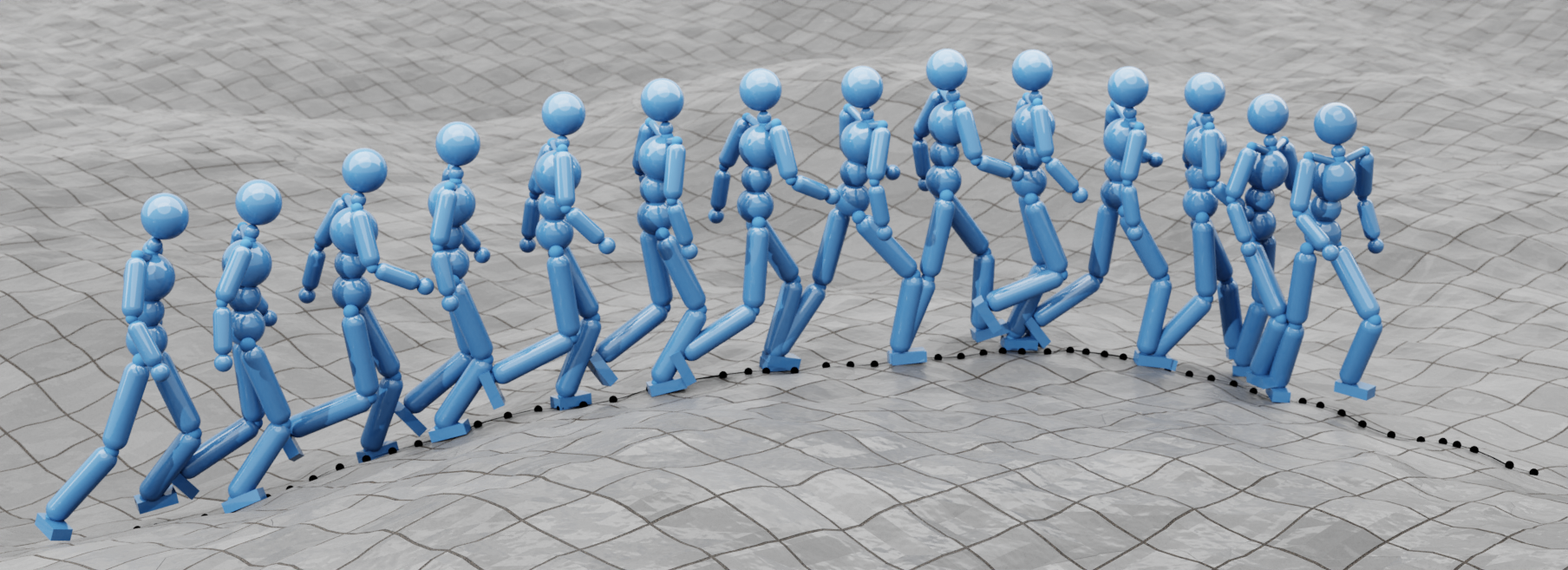}\\\vspace{0.05cm}
    \includegraphics[width=\linewidth]{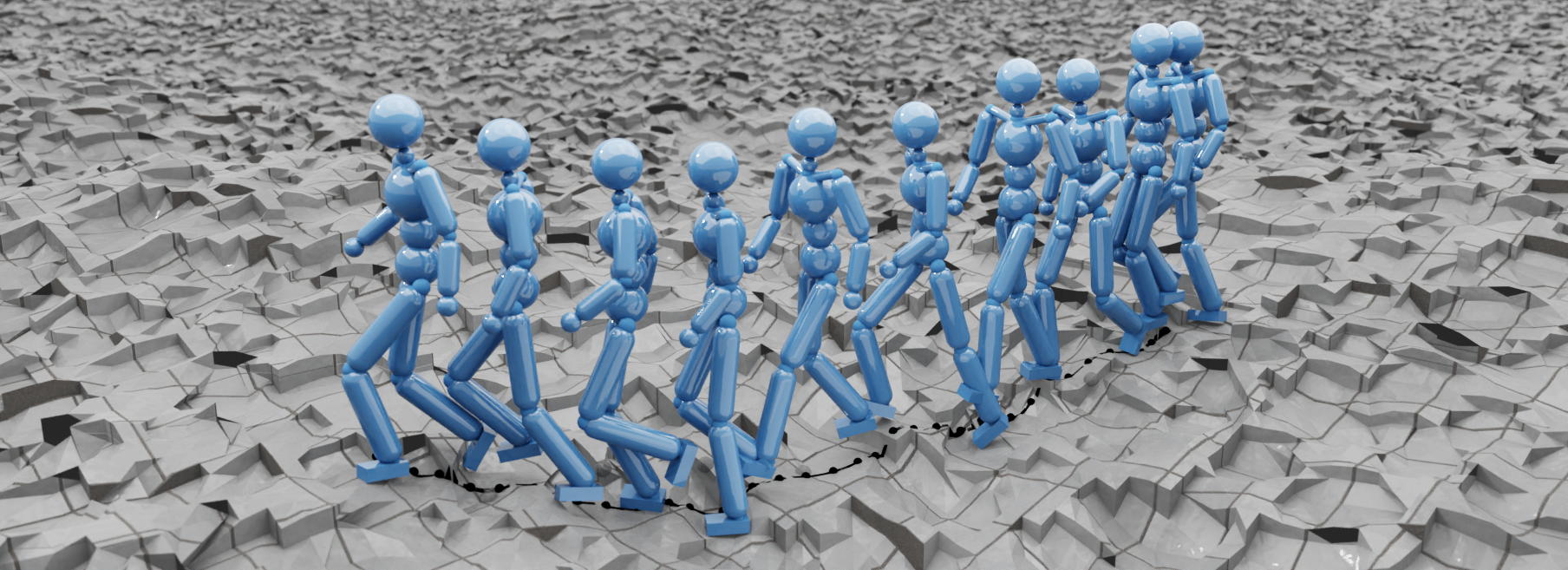}
    \caption{Character controlled with AdaptNet navigates in the environment with procedurally generated terrains. %
    }
    \label{fig:terrain_traj}
\end{figure}

To test AdaptNet with extra control input, 
we designed several experiments where the character is asked to do goal-steering navigation in challenging environments with procedurally generated terrains.
A local heightmap is provided as the additional control input $\mathbf{c}_t$ through which the character is expected to adjust its motions to prevent falling down during walking.
The heightmap is extracted locally based on the character's root position and aligned with the orientation of the root,
with a left and right horizon of 1.7~m, backward horizon of 1~m and forward horizon of 2.4~m.
To process the heightmap $\mathbf{c}_t$, we introduce a convolutional neural network (CNN) as the encoding module $\mathcal{G}_\phi$ (see Eq.~\ref{eq:injection_component}) for AdaptNet.
We refer to the appendix %
for the network structure of the CNN.
An extra map encoding module having the same structure with $\mathcal{G}_\phi$ is added to the critic network for value estimation during adaptation.
We show representative examples %
of our tested terrains in Figure~\ref{fig:terrain_traj} and note the appendix also gives more detail on terrain.

\begin{figure}[t]
    \centering
    \includegraphics[width=\linewidth]{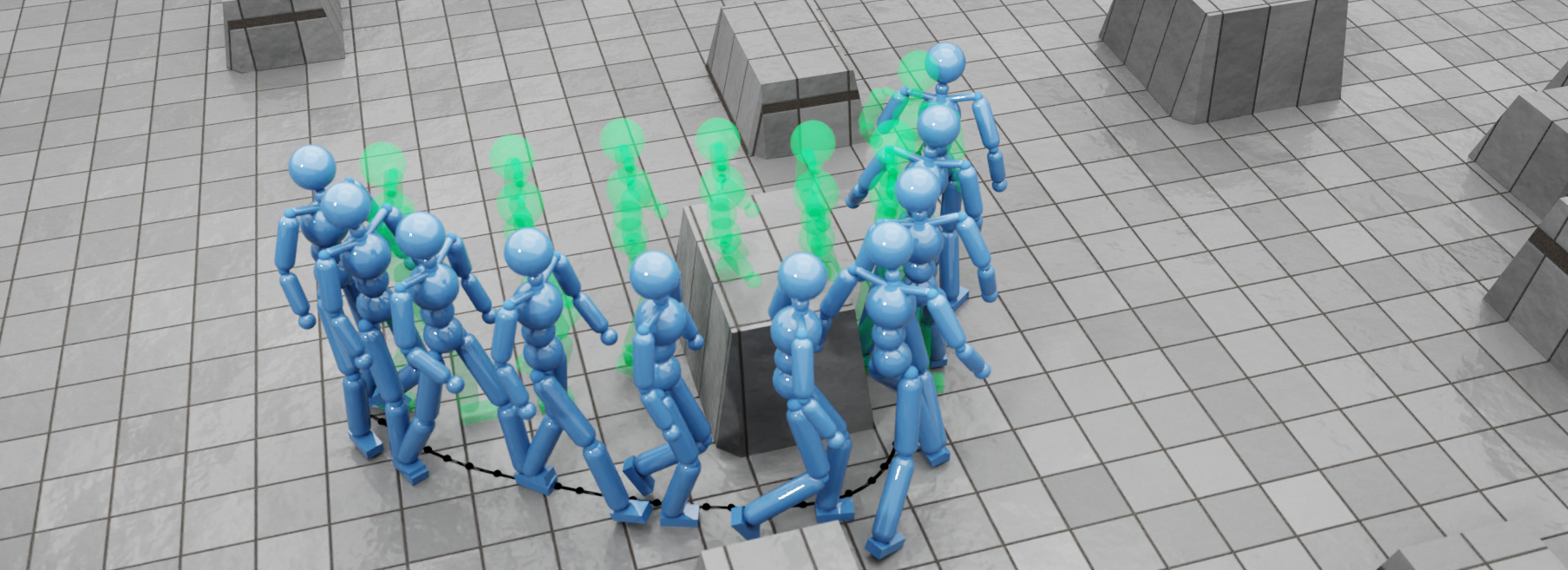}\\\vspace{0.05cm}
    \includegraphics[width=\linewidth]{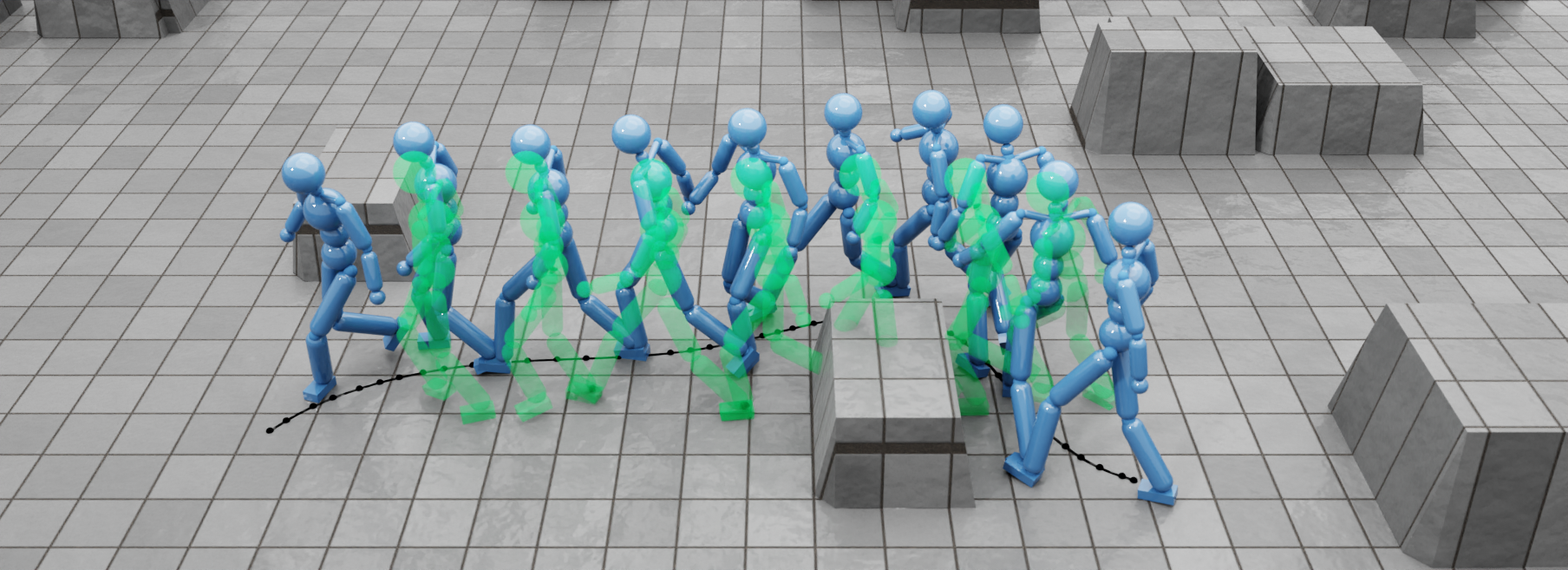}
    \caption{Local collision avoidance in an obstacle-filled environment using AdaptNet. Green characters show the movement trajectory generated by the %
    original walking policy without AdaptNet.  
    }
    \label{fig:terrain_ca}
\end{figure}

We refer to the companion video for the navigation performance of the character when walking on the %
designed terrains after adaptation training.
Even in %
terrains where the height changes smoothly,
the character teeters under the control of the pre-trained policy and a minor change in the terrain slope is enough to make the character stumble.
After adaptation training,
AdaptNet can 
enable the character to smoothly walk and turn on the uneven terrains without falling. 
Besides being able to step over low-height obstacles, the AdaptNet character exhibits intelligent local decision making, trying not 
not to step on the edge of the rocks on the rough terrain and avoids overly rugged paths by altering its moving trajectory to an easy-to-follow one.

To further demonstrate the ability of AdaptNet to perform local path planning,
we designed a more challenging environment with uncrossable obstacles randomly placed on the ground.
We qualitatively show the results in Figure~\ref{fig:terrain_ca}.
As seen in the figure, the character controlled with AdaptNet (blue) can successfully %
walk around the obstacles.
Without accounting for collisions, the character controlled solely by the initially trained policy (green) crosses through the regions where obstacles are placed.

Unsurprisingly, the introduction of the CNN (detailed in Appendix~B) increases the time needed to perform policy optimization iterations in the training for rough terrains.
Still, for the easier terrains, 
training can be done within 1.5 hours. %
The more rugged terrain took around 4 hours for training. 
Finally, it took around 22 hours to train adaptation 
for the local obstacle avoidance test case. 
We note that this is still less time than is needed 
for training the original flat-ground locomotion policy from scratch (26 hours).

\subsection{Perturbation Adaptation}

In a final experimental foray, we investigate AdaptNet's ability to improve the handling of perturbations. Although the original policy can handle small perturbations, the character will still fall under larger impulses. In order to achieve more robust control, we adapt %
the control policy's ability to maintain balance in the presence of large disturbances. We begin with pre-trained policies for target-directed locomotion for walking and running. During the training process, we randomly apply perturbations ($1000$~N, lasting for  $0.2$ seconds) in different directions on the character's torso. With adaptation training of around 5 hours, the character is able to stay balanced against comparable impulses following training for both running and walking tasks.
In contrast, the original controls are not able to handle such perturbations repeatably and they often lead to the characters falling over. Furthermore,  we also observe that AdaptNet control adjusts the character's footsteps to recover balance when the character is highly out of balance due to perturbations.
A comparison of the original policy and our results can be seen in the supplementary video.

\section{Ablation Studies}
\label{sec:comparison}

In this Section, we %
compare 
the performance of AdaptNet %
to different baselines along with performing sensitivity analysis on the two components of the proposed AdaptNet technique.

\subsection{Baseline Comparisons}
\label{sec:baselines}
\begin{figure}[t]
    \centering
    \includegraphics[width=\linewidth]{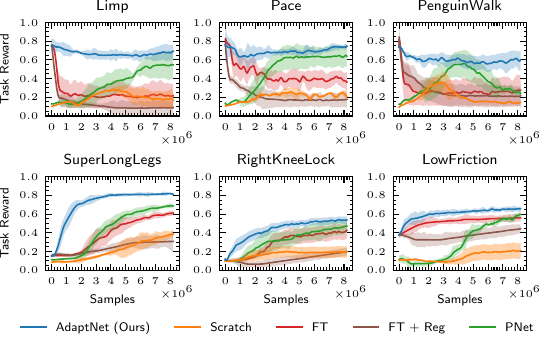}
    \caption{Learning performance of our adaptation scheme using AdaptNet, training from scratch for each task (Scratch), using a progressive network (PNet), and adaptation via directly finetuning the pre-trained policy (FT) and finetuning with regularization (FT + Reg). Colored regions denote mean values $\pm$ a standard deviation based on 5 trials. The top row consists of 
    motion style transfer tasks, while the bottom row focuses on morphological and terrain adaptation tasks.  %
    }
    \label{fig:baseline}
\end{figure}

We consider the following baselines:  \emph{Scratch} where a new policy is trained from scratch on a given task; \emph{FT} where we directly finetune the pre-trained policy network to the newly given task; \emph{FT + Reg} where we apply regularization on the weights of the policy network during finetuning; and \emph{PNet} where policy adaptation is performed using a progressive neural network approach~\cite{rusu2016progressive}.  
Figure~\ref{fig:baseline} compares the 
learning curves for the goal-task performance between  the baselines and AdaptNet 
on three style-transfer tasks (top row) and three adaptation tasks (bottom row), two involving changes in the character's morphology and one for lowered ground friction.  
For fair comparison, we employ the same 
training setup 
for all baselines, where the reward function of the new policy accounts for both a task objective and an imitation objective using an automatic weighting scheme~\cite{composite}. 
In the motion style transfer experiments, the imitation term is computed using a new discriminator that takes only the stylized motions as the reference similar to Section~\ref{sec:style_transfer}.

\begin{figure}[t]
    \centering
    \includegraphics[width=\linewidth]{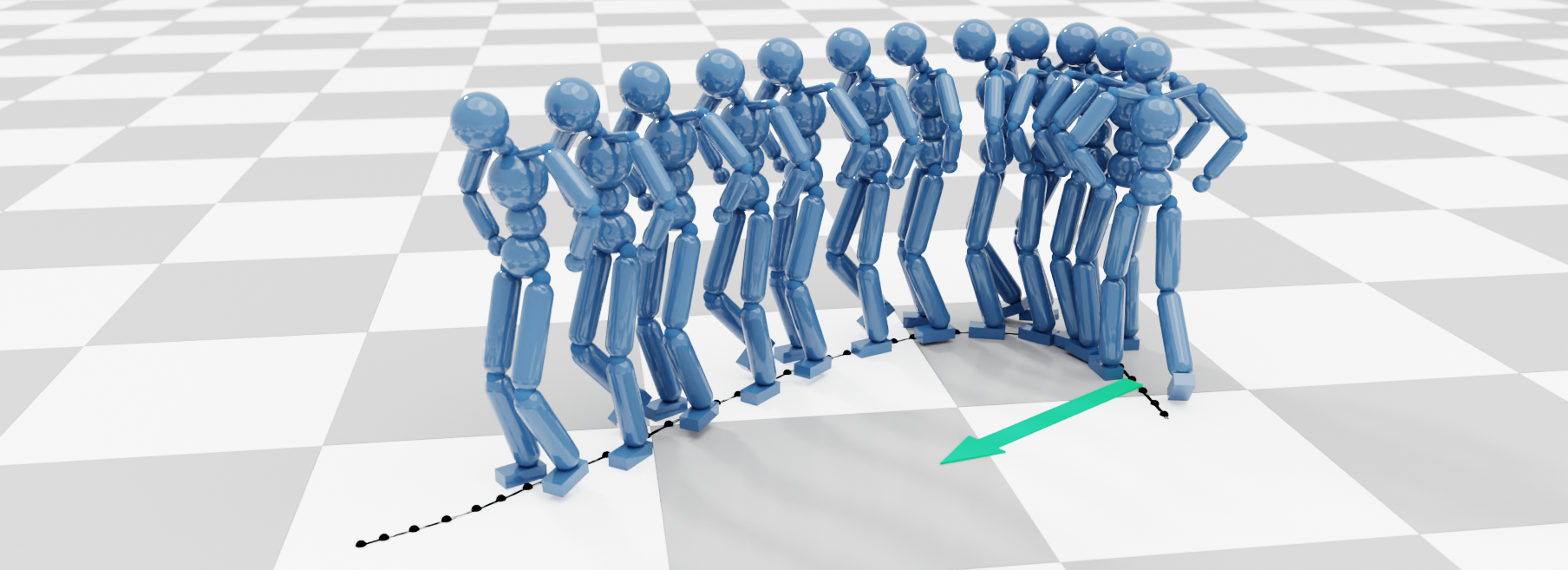}\\\vspace{0.05cm}
    \includegraphics[width=\linewidth]{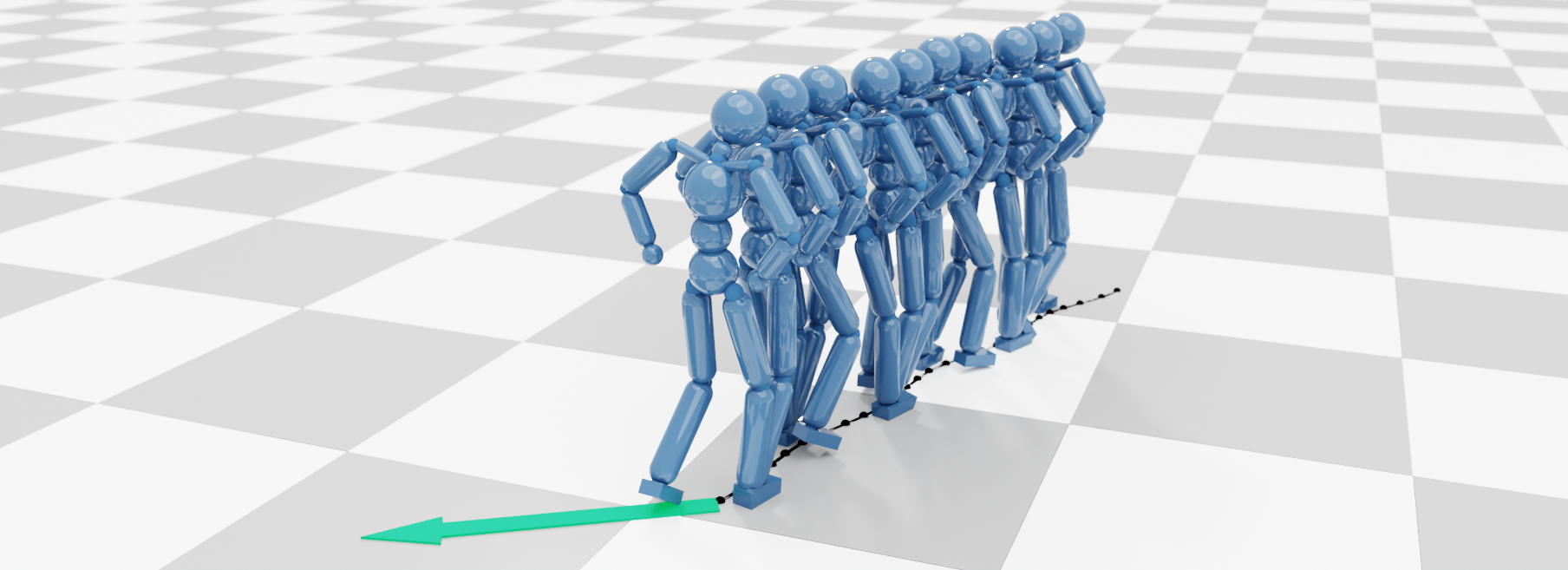}
    \caption{Top: AdaptNet successfully controls the character to turn during walking in \textit{Pace} style. Bottom: The character controlled by FT policy keeps imitating the reference motion to pace straightly, and fails to turn due to overfitting. Green arrows indicate the dynamically generated target directions for locomotion control.}
    \label{fig:baseline_fail}
\end{figure}

As can be seen from the learning curves in Figure~\ref{fig:baseline}, %
\emph{Scratch} 
fails to attain the desired goals in the considered benchmarks, achieving a very low goal task reward within the given budget of 8M training samples.
\emph{FT} can effectively modify the locomotion policy in the bottom three tasks where the character's morphology or environmental friction changes. %
However, in the motion style transfer tasks, the reward curve of \emph{FT} noticeably drops after the training begins
as
\emph{FT} overfits the imitation of the newly provided stylized reference motion and ignores the goal direction signal.  %
In contrast, AdaptNet provides a stable task reward curve during the adaptation training with the character being able to imitate the newly provided style without forgetting the previously learned locomotion behaviors as seen in  Figure~\ref{fig:baseline_fail}. 
The above findings are in line with previous works~\cite{MCPPeng19,rusu2016progressive} that have shown finetuning to be efficient when the parameters of a pre-trained model need to be slightly adjusted to a new target domain.  
However, \emph{FT} can be susceptible to catastrophic forgetting
when the imitation objective is significantly changed, as in the motion style transfer tasks.
\emph{FT + Reg} leads to poor training and low-fidelity controllers in all tasks. While, in theory, adding regularization can improve the navigation performance, in practice, it is hard to regulate the weights during finetuning due to the presence of both significant large and small weights in the pre-trained policy.   %

\emph{PNet} shares similarities with AdaptNet as both approaches add new weights to the original policy 
network and freeze the old weights during transfer learning. However, despite these similarities, the architectures of the two approaches are significantly different. AdaptNet uses a residual structure that supports merging, resulting in a single policy network which allows forward propagation in one pass during inference. In contrast, \emph{PNet} does not support merging and requires the original network to be present and run first to compute the values of the hidden neurons in the added network. 
This adds significant complexity and memory overhead, with the network structure becoming larger and slower.
Importantly, during training, the added network in \emph{PNet} cannot start from zero as compared to AdaptNet. In essence, the zero initialization in AdaptNet allows us to guide the adaptation starting from the original policy.  This is clear in the
style-transfer tasks, where AdaptNet begins training with a much higher reward than \emph{PPNet} due to the locomotion ability provided by the original policy. 
Despite its competitive final performance in several of the adaptation tasks, \emph{PNet} is sample inefficient. %
Finally, we note that %
it can lead to forgetting the prior knowledge provided by the pre-trained policy as the added network 
can
significantly change the output of the whole model in some cases. 
This can be seen in the \textit{Penguin Walk} task where the navigation performance drops after 5M samples.

Overall, AdaptNet consistently outperforms all four baselines in terms of final performance and sample efficiency. In terms of memory efficiency, \emph{Scratch} and \emph{FT} do not add any overhead.
AdaptNet introduces additional parameters, but since 
the original network is frozen, 
the number of trainable parameters is still at the same scale with the original neural network when no conditional input, i.e., $c_t$ and $\mathcal{G}_\phi$, is needed.  
While the the total number of parameters 
increases, the effective number of parameters is the 
same as the original policy because
AdaptNet can be merged into the original network. In contrast, \emph{PNet} requires both networks to be present and effectively doubles the number of parameters.

\subsection{Latent Space Injection}\label{sec:exp_latent_inject}
\begin{figure}[t]
\centering
\includegraphics[width=0.23\linewidth]{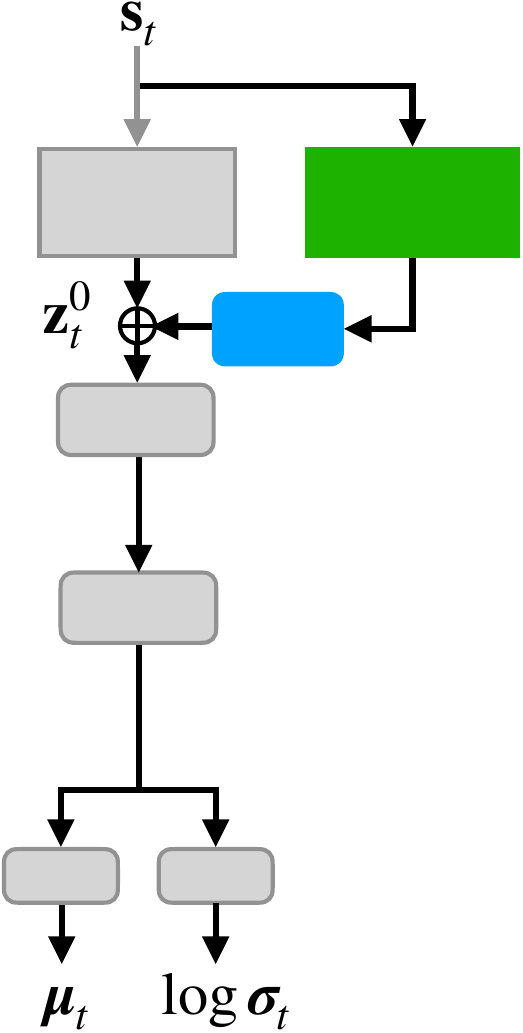}\hfill
\includegraphics[width=0.23\linewidth]{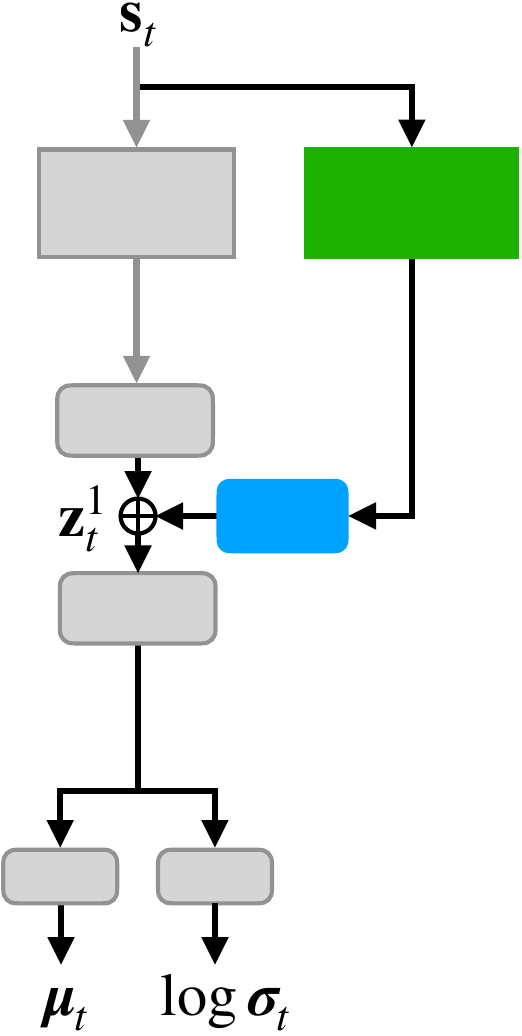}\hfill
\includegraphics[width=0.23\linewidth]{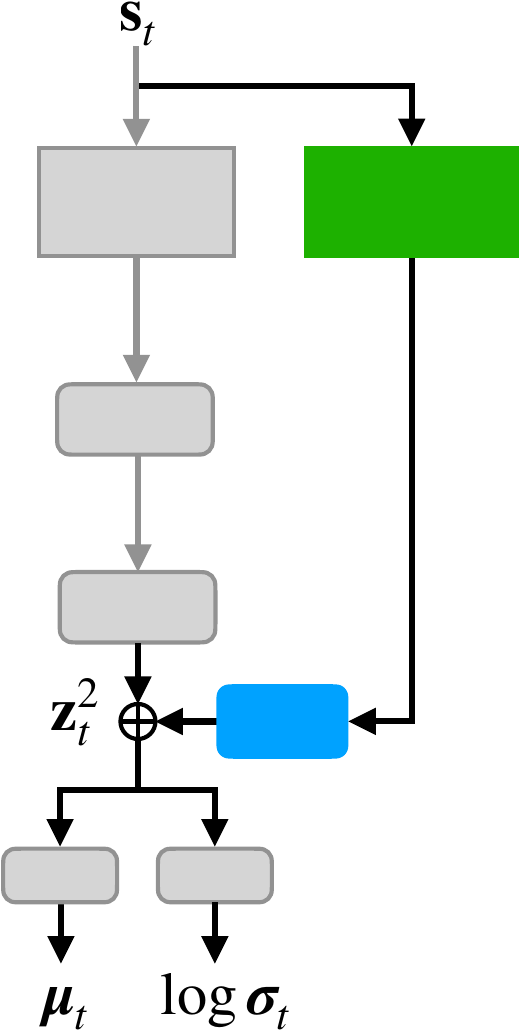}\hfill
\includegraphics[width=0.23\linewidth]{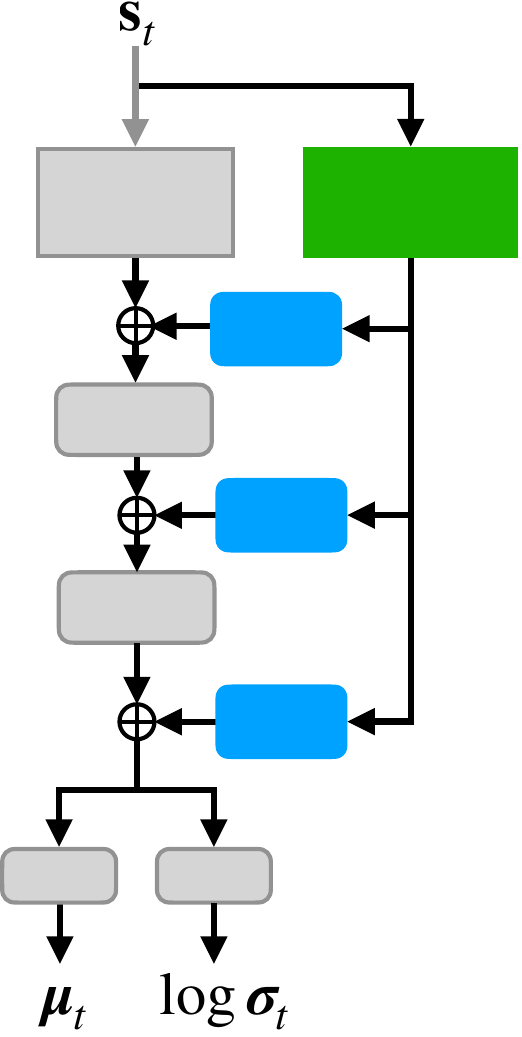}
    \caption{Injection at different latent spaces. Gray blocks represent the original policy network locked during adaptation. Green blocks are the state encoder $\mathcal{E}_\phi$, and blue ones are $\mathcal{F}_\phi$.
    For left tor right, the manipulated latent spaces are $\mathcal{Z}^0$ (the default implementation of AdaptNet), $\mathcal{Z}^1$, $\mathcal{Z}^2$ and $\mathcal{Z}^{0:2}$ respectively.
    We ignore $\mathcal{G}_\phi$, given that there is no extra control input in the tested examples here.} 
    \label{fig:network_inject_space}
\end{figure}

Our default implementation performs %
injection on the latent space $\mathcal{Z}^0$ right after the goal state $\mathbf{g}_t$ and character state $\mathbf{o}_t$ are encoded and concatenated together. 
Here, we test the application of the injection module to other latent spaces after $\mathcal{Z}^0$ but before reaching the action space, along with applying injection on all possible latent spaces simultaneously. 
To solely study the performance of latent space injection,
we also remove the full-rank adaptation modules 
for these tests.
The tested network structures are shown in Figure~\ref{fig:network_inject_space}.

\begin{figure}[t]
    \centering
    \includegraphics[width=\linewidth]{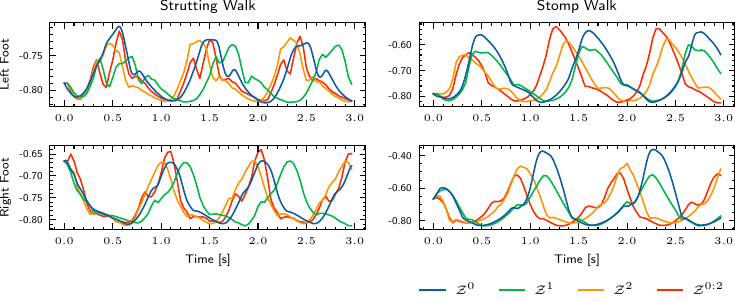}
    \caption{Foot height (in meters) relative to the root when performing adaptation for motion style transfer tasks with different latent spaces being injected. Injection at $\mathcal{Z}^0$ (blue) leads to the smoothest and most repeatable stepping motions.}
    \label{fig:latent_inject_pos}
\end{figure}

To explore how the injection schemes perform 
differently in generating new policies,
we run tests on several motion style transfer tasks. %
During our experiments, we observe qualitatively that 
injection at the lower space $\mathcal{Z}^2$ or at all the latent spaces $\mathcal{Z}^{0:2}$,   which also includes the lower one, 
can easily 
produce jerky motions with stiff movements of the torso and legs. 
It can also lead to failures in training where the character falls repeatedly after a few training iterations. %
In Figure~\ref{fig:latent_inject_pos}, we plot the trajectory of the foot height in two of our tested cases.
While injection at $\mathcal{Z}^0$ (blue) leads to a smooth repeatable trajectory, %
the curves become more irregular as the injected latent space changes from $\mathcal{Z}^1$ (green) to $\mathcal{Z}^2$ (orange) and then to $\mathcal{Z}^{0:2}$ (red). 
We also see some sharp jumps in the curves of $\mathcal{Z}^2$ and $\mathcal{Z}^{0:2}$,
which represent fast motion transitions.
We refer to the supplementary video for the animation results including examples where injection at $\mathcal{Z}^2$ and $\mathcal{Z}^{0:2}$ fails.

Overall, our tests show that as the chosen target latent space is closer to the action space, it becomes more difficult for AdaptNet to generate desired motions, with $Z^0$ both intuitively and empirically giving the best results. 
This observation is in agreement with recent work in  
image synthesis where the target space for manipulation is usually chosen nearer to the input of the generator rather than near the final output~\cite{abdal2019image2stylegan,zhuang2021enjoy,karras2020analyzing}. 
In terms of the network structure in our implementation, 
the input state $\mathbf{s}_t \in \mathbb{R}^{784}$ is encoded into the first latent space $\mathcal{Z}^0 \in \mathbb{R}^{260}$ and then projected to $\mathcal{Z}^1 \in \mathbb{R}^{1024}$.
The whole network, therefore, can be regarded as an encoder-decoder structure where the bottleneck is at $\mathcal{Z}^0$.
As we will show in Section~\ref{sec:latent}, $\mathcal{Z}^0$ is well-structured 
which makes it amenable to manipulation %
for motion generation. 

\subsection{Comparison of Adaptation Methods}\label{sec:lora}

We quantitatively evaluate the imitation performance of AdaptNet
with other adaptation approaches, including alternate methods 
with and without using its internal adaptation component.  
As in prior work~\cite{harada2004quantitative,tang2008emulating,peng2021amp,iccgan},
we 
measure the imitation error via:
\begin{equation}
    e_t = \frac{1}{N_\text{link}} \sum_{l=1}^{N_\text{link}} \| p_l - \tilde{p}_l\|, 
\end{equation}
where 
$N_\text{link}$ is the total number of body links,
$p_l \in \mathbb{R}^3$ is the position of the body link $l$ in the world space at the time step $t$, and $\tilde{p}_l$ is the body link's position in the reference motion. 
The evaluation results are shown in Table~\ref{tab:model_adaptation}.

\begin{figure*}[t]
    \centering
    \includegraphics[width=\linewidth]{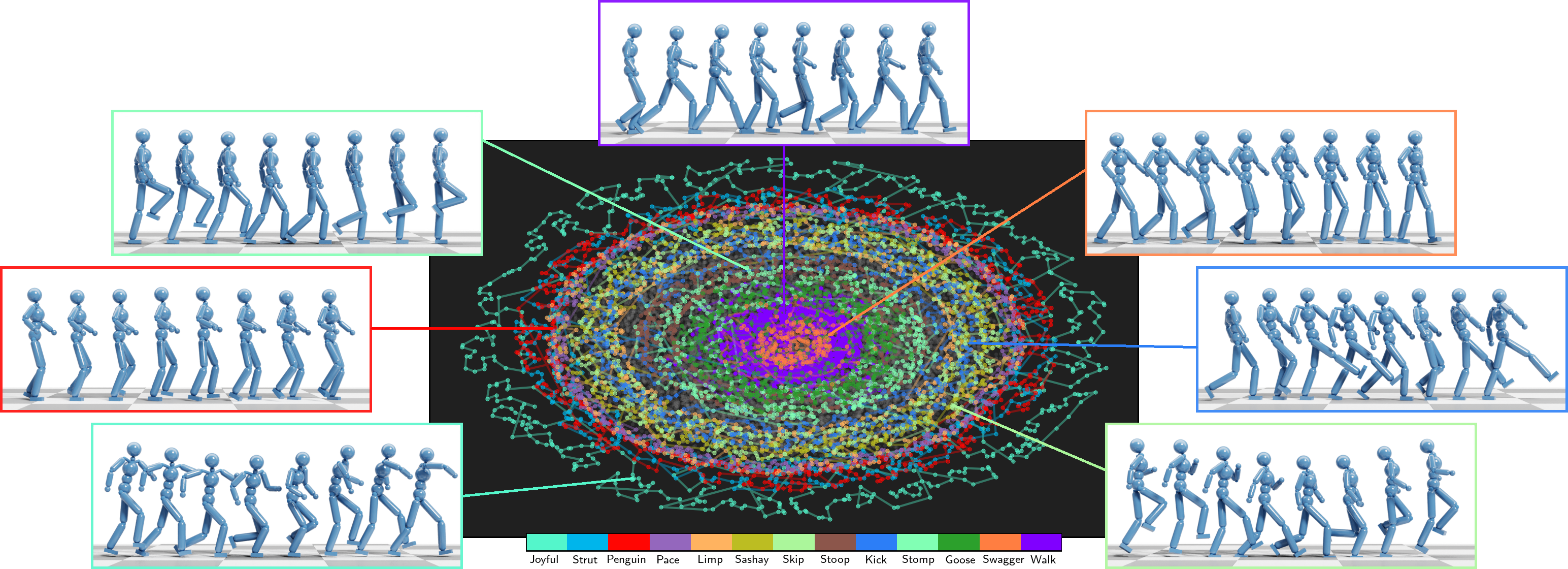}
    \caption{Latent space visualization with respect to different styles of walk-related motions. The latent representations of the stylized motions are obtained by AdaptNet without using the internal adaptation component. The walk motion (dark purple) near the center is provided by the pre-trained policy based on which AdaptNet performs adaptation to learn the stylized motions. The visualization is achieved using multidimensional scaling technique to project the latent representations from 260 dimensions to 2 dimensions. } \vspace{-2pt}
    \label{fig:latent}
\end{figure*} %

We find our proposed approach to combine latent space adaptation (LSA) and internal adaptation (IA) results in the best performance. %
While the results in Table~\ref{tab:model_adaptation} imply LSA alone is sufficient in many cases, 
IA appears to help most in the difficult motion style transfer tasks, e.g., \textit{Goose Step}, \textit{Jaunty Skip} and \textit{Joyful Walk}, where the stylized motions are relatively far away from the pre-trained walking motions. 
In these tasks, adding 
IA improves the visual quality as well as motion smoothness, foot height, and gait frequency as shown in the supplementary video. 
It is important to note, however, that IA alone produces subpar performance.
In addition, it cannot account for the additional control input needed in other adaptation tasks such as terrain adaptation. 
Further, even when no additional input is needed, IA components cannot be applied for modification of the state encoder as we cannot initialize the GRU layer of the encoder to zero. Latent modification is a distinctive feature of LSA,  %
rendering Eqs.~\ref{eq:latent_after_injection} and~\ref{eq:adapt_component} unsuitable to be merged into the same formulation.
To highlight the importance of the state encoder module $\mathcal{E}_\phi$ in LSA, we consider an additional ablation 
where we remove $\mathcal{E}_\phi$ and connect the output of $\mathcal{E}_\xi$ directly to $\mathcal{F}_\phi$ (see Figure~\ref{fig:overview}). 
As shown in Table~\ref{tab:model_adaptation}, utilizing just the old latent space embedding is useful but no more valuable than using internal adaptation.

In addition to %
ablations to our own architecture, 
we also compare our IA component, which can be regarded as a full-rank adaptation scheme, to the low-rank adaptation (LoRA) scheme~\cite{hu2021lora}. 
LoRA typically works well for adapting large language models with a low rank $\leq$ 8.
However, we did not find any evident improvement over just using LSA when an intrinsic rank of 8 was employed.
Even after increasing the rank to 64,
the performance gap between the full-rank adaptation scheme and LoRA still remains as listed in Table~\ref{tab:model_adaptation}. %
Though using a low-rank decomposition can reduce the total number of parameters,
it increases the computation cost since one more matrix multiplication is needed for each adaptor.
Given the small size of our policy network,  %
from our findings we conclude that the full-rank adaptation offers desirable benefits over LoRA. 

\begin{table}[t]
    \setlength\tabcolsep{0.085cm}
    \centering\small
    \caption{%
    Imitation error during motion style transfer with different adaptation components.
    Values are reported in meters in the format of mean$\pm$std.}\vspace{-6pt}
    \begin{tabular}{r|c|c|c|>{}c|>{}c}
        \toprule
        \multirow{2}*{\textbf{Motion}} & \textbf{AdaptNet} & \textbf{LSA}  & \multirow{2}*{\textbf{LSA}}& \textbf{LSA} & \multirow{2}*{\textbf{IA}} \\
         & \textbf{(LSA+IA)} & \textbf{+LoRA-64} & & \textbf{w/o $\mathbf{\mathcal{E}_\phi}$} &  \\
        \midrule
        \footnotesize{Swaggering} & $\mathbf{0.05\pm0.02}$ & $\mathbf{0.05\pm0.02}$ & $0.06\pm0.02$ & $0.11\pm0.03$ & $0.11\pm0.03$ \\
        \footnotesize{Goose Step} & $\mathbf{0.11\pm0.08}$ & $0.18\pm0.08$ & $0.21\pm0.12$ & $0.35\pm0.11$ & $0.36\pm0.11$ \\
        \footnotesize{Stomp} & $\mathbf{0.08\pm0.04}$ & $0.10\pm0.05$ & $0.11\pm0.06$ & $0.26\pm0.07$ & $0.27\pm0.08$   \\
        \footnotesize{Kicking} & $\mathbf{0.08\pm0.03}$ & $\mathbf{0.08\pm0.03}$ & $0.09\pm0.05$ & $0.20\pm0.07$  & $0.21\pm0.07$ \\ %
        \footnotesize{Stoop} & $\mathbf{0.07\pm0.02}$ & $\mathbf{0.07\pm0.02}$ & $\mathbf{0.07\pm0.02}$ & $0.14\pm0.03$ & $0.13\pm0.03$ \\
        \footnotesize{Jaunty Skip} & $\mathbf{0.16\pm0.09}$ & $0.22\pm0.10$ & $0.25\pm0.12$ & $0.56\pm0.18$  & $0.61\pm0.21$\\
        \footnotesize{Sashay} & $\mathbf{0.06\pm0.03}$ & $\mathbf{0.06\pm0.03}$ & $\mathbf{0.06\pm0.03}$ & $0.09\pm0.03$ & $0.09\pm0.04$\\
        \footnotesize{Limp} & $\mathbf{0.10\pm0.07}$ & $\mathbf{0.10\pm0.07}$  & $0.12\pm0.07$ & $0.22\pm0.09$ & $0.29\pm0.11$ \\
        \footnotesize{Pace} & $\mathbf{0.09\pm0.03}$ & $0.10\pm0.03$ & $0.10\pm0.03$ & $0.14\pm0.03$ & $0.13\pm0.03$ \\
        \footnotesize{Penguin} & $\mathbf{0.11\pm0.04}$ & $0.13\pm0.05$ & $0.15\pm0.05$ & $0.31\pm0.09$ & $0.38\pm0.13$ \\ %
        \footnotesize{Strutting} & $\mathbf{0.09\pm0.03}$ & $0.10\pm0.05$ & $0.12\pm0.06$ & $0.23\pm0.06$ & $0.27\pm0.06$ \\ %
        \footnotesize{Joyful} & $\mathbf{0.17\pm0.07}$ & $0.22\pm0.09$ & $0.28\pm0.12$ & $0.54\pm0.22$ & $0.59\pm0.24$ \\
        \bottomrule
    \end{tabular}
    \label{tab:model_adaptation}
\end{table}

\section{Latent Space Analysis}
\label{sec:latent}

In this section, we provide more insights on the ability of AdaptNet to successfully control and modify the latent space.  
\subsection{Latent Space Visualization}

Figure~\ref{fig:latent} visualizes the latent space for different motion style transfer tasks. %
For each task, a controller was trained using AdaptNet starting from the same pre-trained locomotion policy of walking. 
During adaptation training here, we use only the latent space injection component 
as in $\mathcal{Z}^0$ for all  models. %
We also remove the regularization term $\mathcal{I}_\phi$ in Eq.~\ref{eq:loss_adapt} and prolong the training time to
let AdaptNet fit the style motions as much as possible. 
After training,
we collect samples for each stylized motion from the simulated character following  a straight path without any goal-direction changes.  
We use a multidimensional scaling technique to reduce the dimension of the collected latent samples.

As seen in the figure, the 2D projection of the latent space 
exhibits a circular shape with the pre-trained walking policy (dark purple) located near the center. 
There is a clear and roughly continuous transition when the motion style changes from one to the other,
which demonstrates the well structured nature of the latent space %
with the different motion styles.
The distribution of the stylized motion in the visualized %
space 
is roughly consistent with the imitation error distribution listed in Table~\ref{tab:model_adaptation} when no 
internal adaptor is employed.
Motions with smaller imitation errors are distributed generally closer to the pre-trained policy while
\textit{Joyful Walk} (light green) has the largest error and is located the farthest away from the center of the circle. 
We also note the \textit{Penguin Walk} (red) and \textit{Pace} (light purple) show greater differences in frequency and speed and %
appear farther away from the center of the figure. 
This indicates that the distribution in the latent space not only reflects the pose similarity between motions but also some semantic information, like motion rhythm and gait frequency.
Similar conclusions have been drawn by recent work %
in the field of image generation, where the latent space for image generation is considered to capture semantic information more than just simple color transformations~\cite{Jahanian2020On,epstein2022blobgan,shen2020interpreting}.  
\begin{figure*}[t]
    \centering
    \includegraphics[width=.23\linewidth]{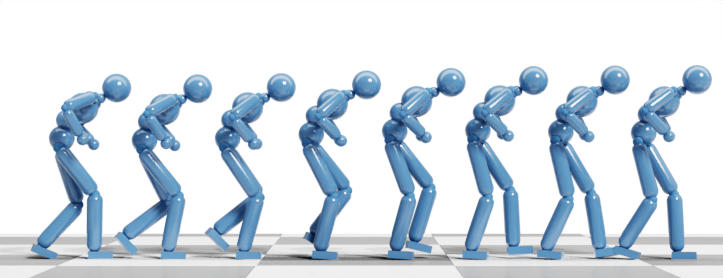}
    \quad
    \includegraphics[width=.23\linewidth]{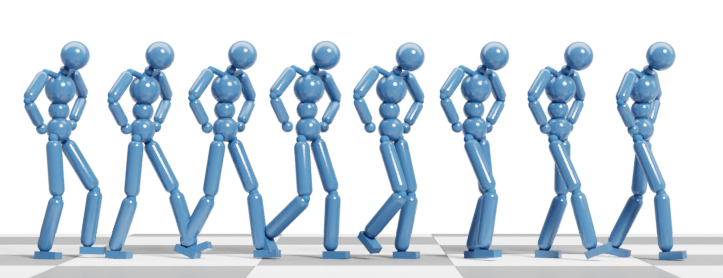}
    \quad
    \includegraphics[width=.23\linewidth]{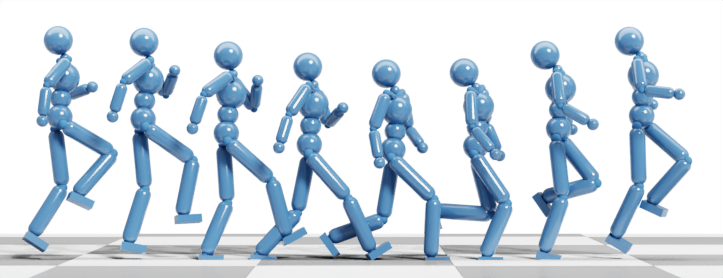}
    \quad
    \includegraphics[width=.23\linewidth]{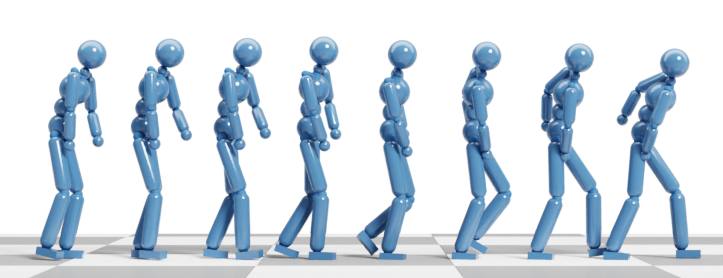}
    \includegraphics[width=.23\linewidth]{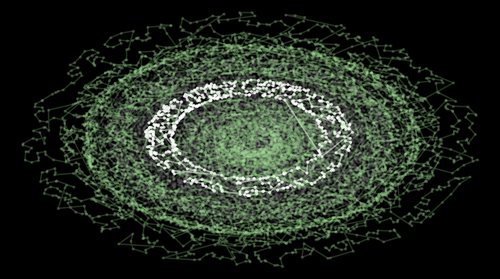}
    \quad
    \includegraphics[width=.23\linewidth]{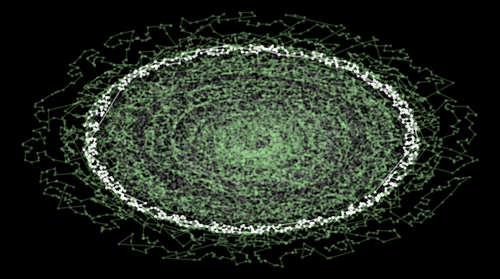}
    \quad
    \includegraphics[width=.23\linewidth]{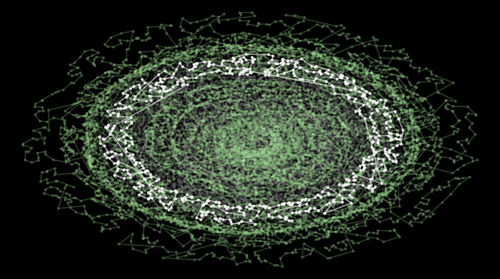}
    \quad
    \includegraphics[width=.23\linewidth]{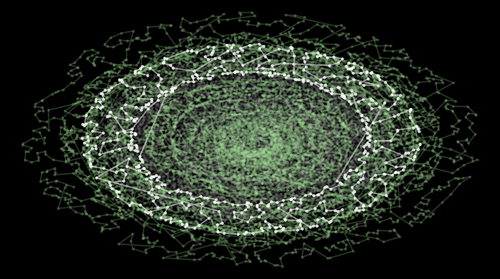}
    \includegraphics[width=.23\linewidth]{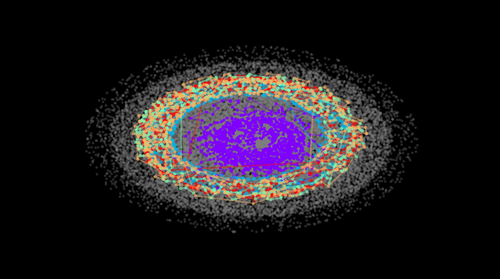}
    \quad
    \includegraphics[width=.23\linewidth]{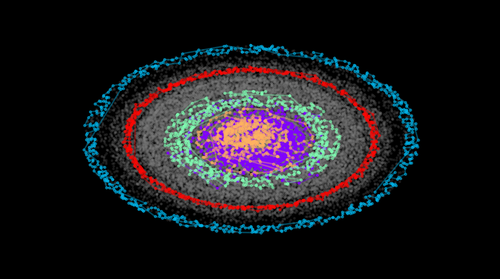}
    \quad
    \includegraphics[width=.23\linewidth]{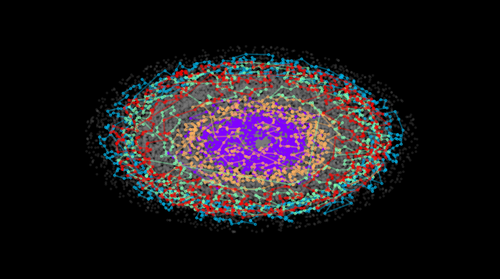}
    \quad
    \includegraphics[width=.23\linewidth]{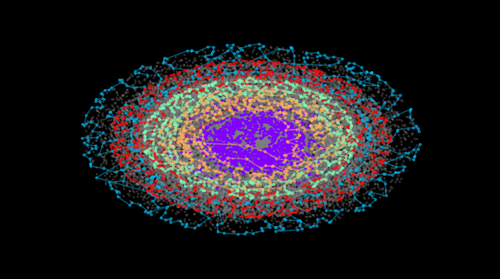}
    \\\vspace{0.2cm}
    \includegraphics[width=.65\linewidth]{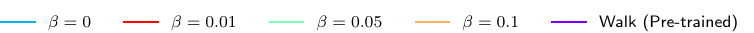}

    \vspace{-8pt}
    \caption{Latent space distributions of \textit{Stoop},  \textit{Pace}, \textit{Jaunty Walk},  \textit{Limp} (left to right). 
    The top figures show the distribution of the stylized motions in the full latent space without regularization, and
    the bottom figures show the distribution with regularization applied during adaptation training.
    Gray points shown in the bottom figures are the latent embeddings generated by the pre-trained encoder $\mathcal{E}_\xi$ while the character performs the stylized motions.
    $\beta$ is the regularization coefficient from Eq.~\ref{eq:loss_adapt}.
    }
    \label{fig:latent_l2}
\end{figure*}

\subsection{Latent Injection Regularization}
In Figure~\ref{fig:latent_l2}, 
we show the latent visualizations of several motions %
generated by AdaptNet when L2 regularization is applied on the injected latent.  
For comparison, we highlight in white each motion's distributions %
in the full latent space shown in Figure~\ref{fig:latent}.
In the lower figures,
the dark purple points represent the latent embedding of the pre-trained walking, while %
the gray points are generated by the pre-trained encoder $\mathcal{E}_\xi$ when the simulated character performs stylized motions. %
Other colors represent varying levels of regularization, as shown.
The goal of regularization is to ensure that the generated latent can fall into the manifold composed of the gray dots. 
This represents a relatively safe region 
where the latent space is expected to be handled properly by %
the pre-trained policy.

In the \textit{Stoop} task, there is almost no difference with and without using the L2 regularization.
All visualized samples are overlapped together and covered by the gray region. This is expected 
given that the style motion of \textit{Stoop} is close to the walking motion in the latent space.
However, in the example of \textit{Pace}, there is a clear separation when different regularization coefficients are employed.
Note when a coefficient of 0.1 is taken, the generated stylized motion (orange) is overlapped with the walking motion (dark purple).
AdaptNet, in this case, is over-regularized.
It yields to the pre-trained policy and fails to adapt the pre-trained policy to perform the desired stylized motion.
In contrast, without regularization ($\beta=0$), the latent is already outside of the safe, gray region. 
AdaptNet, in this case, simply overfits to imitating the style motion and loses the ability to perform goal-steering navigation. 
While in \emph{Jaunty Skip}, any $\beta$-value can be employed, in \emph{Limp} a $\beta$-value of 0.01 best ensures that the latent space stays into the grey manifold while attaining high imitation performance. In all adaptation tasks detailed in the paper, we found $\beta = 0.01$ to be sufficient.
We note that such regularization is not necessary in other tested adaptation tasks without motion style transfer. In such cases, the new expected motions are close to the original policy and already lie in the safe region.
We refer 
to the supplementary video for a visual comparison of the generated motions when different regularization coefficients are employed. %

\section{Conclusions} 
\label{sec:conclusions}

This paper presents AdaptNet, an approach for adapting existing character RL-based control policies to new motion tasks.  Our approach applies two strategies. The first adapts the latent space by conditioning on the character's state and allowing the addition of new control inputs that will allow the control policy to perform new tasks. The second aims at control refinement which allows policy adaptation %
by shifting the original policy and generating new control actions based on new training.  
Importantly, AdaptNet training always begins 
with
having no (zero) influence, starting from the existing policy and increasing its %
influence as training proceeds. 

We demonstrate that a previously trained control policy for %
locomotion can be adapted to support diverse style transfer, 
morphological changes including limb length variation and locked joints, and terrain adaptation including varied friction and geometry. %
These adaptations are also very efficient to learn. While the original locomotion policy requires 26 hours of training, our style adaptations take less than thirty minutes to produce a full controller that is capable of goal-directed steering while adhering to a specified walking style. %
More extreme adaptations require more time, but training is still far more efficient than the cost of learning the initial policy.

A core limitation of this work is that policy adaptation requires an existing pre-trained policy, and thus it cannot act to produce new motions on its own.  While it is capable of migrating the policy to many new behaviors and conditions, extreme adaptions (e.g., training a jumping action with long flight phase from a walking controller) do not produce the expected results.  We believe this is due to the distinct characteristics of the two behaviors and we see such `deep' adaptation as a direction for future work.  Also, while we demonstrate smooth interpolation between latent space embeddings when we employ control-layer refinement, interpolation does not always produce coherent in-between behaviors. 
As we show in Section~\ref{sec:exp_latent_inject},
an improper choice of the target latent space could lead to undesired control results. %
As such, we found starting with a proper latent space is important for obtaining high-quality controllers. 

In the current work, we use the recent approach of~\citet{composite} for pre-training an initial policy that is then modified by AdaptNet. In the future, we would like to see how well other recent approaches for training physics-based controllers~\cite{peng2022ase,peng2021amp,Yao2022ControlVAE} can work with our proposed approach. 
We would also like to investigate how our approach can be extended to generate a well-represented latent space that can be further exploited for motion synthesis. This opens up many avenues for further research, including latent space disentanglement, inversion, and shaping.

\begin{acks}
We acknowledge the support of the Natural Sciences and Engineering Research Council of Canada (NSERC) and the National Science Foundation under Grants No. IIS-2047632 and IIS-2232066.  Support for the first author was made through a generous gift from Roblox.  The Bellairs Workshop on Computer Animation was instrumental in the conception of the research presented in this paper.   
\end{acks}

\bibliographystyle{ACM-Reference-Format}
\bibliography{adaptNet}

\appendix

\vspace*{6pt}
\section{Locomotion Task Setup}
The goal state $\mathbf{g}_t \in \mathbb{R}^4$ includes a 2D unit vector representing the direction to the target location, the horizontal distance from the character to the goal, i.e., $\|\mathbf{x}_t^\text{root} - \mathbf{p}_\text{goal}\|$, and the preferred speed $\| \mathbf{v}_t^\ast \|$.
The preferred speed is sampled from $[1, 1.5]$ in the unit of $m/s$ for crouching and walking motions, and from $[1, 3]$ for running.
The goal direction is sampled from $[0, 2\pi)$.
A timer variable is sampled from $[3, 5]$ in the unit of $s$ for walking motions, and from $[2, 3]$ for running.
We use these three goal variables to obtain the target location.
As such, we can perform speed control during the location targeting.
The goal-directed task reward is
\begin{equation*} %
    r_t = \begin{cases}
        \exp(-3\|\dot{\mathbf{x}}_{t+1}^\text{root}/T - \mathbf{v}_t^\ast\|^2/\| \mathbf{v}_t^\ast \|^2) & \text{if } \|\mathbf{x}_{t+1} - \mathbf{p}_\text{goal}\| > R\\
        1 & \text{otherwise},
    \end{cases}
\end{equation*}
where $R=0.5$ is the goal radius of the target location, $T = 1/30$~s is the time interval between two frames,
$\dot{\mathbf{x}}_{t+1}^\text{root}/T$ denotes the horizontal velocity of the root link at time $t+1$, 
and $\mathbf{v}_t^\ast$ is the target velocity toward the goal location.

\section{Hyperparameters}
\begin{table}[t]
\centering\vspace{20pt}
\caption{Training hyperparameters.}\vspace{-12pt}
\begin{tabular}{lc}
    \toprule
    \textbf{Parameter} & \textbf{Value}\\
    \midrule
    policy network learning rate & $5 \times 10^{-6}$\\
    critic network learning rate & $1 \times 10^{-4}$\\
    discriminator learning rate & $1 \times 10^{-5}$\\
    reward discount factor ($\gamma$) & $0.95$ \\
    GAE discount factor ($\lambda$) & $0.95$ \\
    surrogate clip range ($\epsilon$) & $0.2$ \\
    gradient penalty coefficient ($\lambda^{GP}$) & $10$ \\
    number of PPO workers (simulation instances) & $512$ \\
    PPO replay buffer size & $4096$ \\
    PPO batch size & $256$ \\
    PPO optimization epochs & $5$ \\
    discriminator replay buffer size & $8192$ \\
    discriminator batch size & $512$ \\
    Regularizer for internal adaptors ($\beta$) & $0.01$ \\
    Regularizer coefficient for latent space adaptor ($\kappa$) & $0.01$ \\
  \bottomrule
\end{tabular}
\label{tab:hyper}
\end{table}

\begin{figure}[t]
    \centering\vspace{-4pt}
    \includegraphics[width=0.96\linewidth]{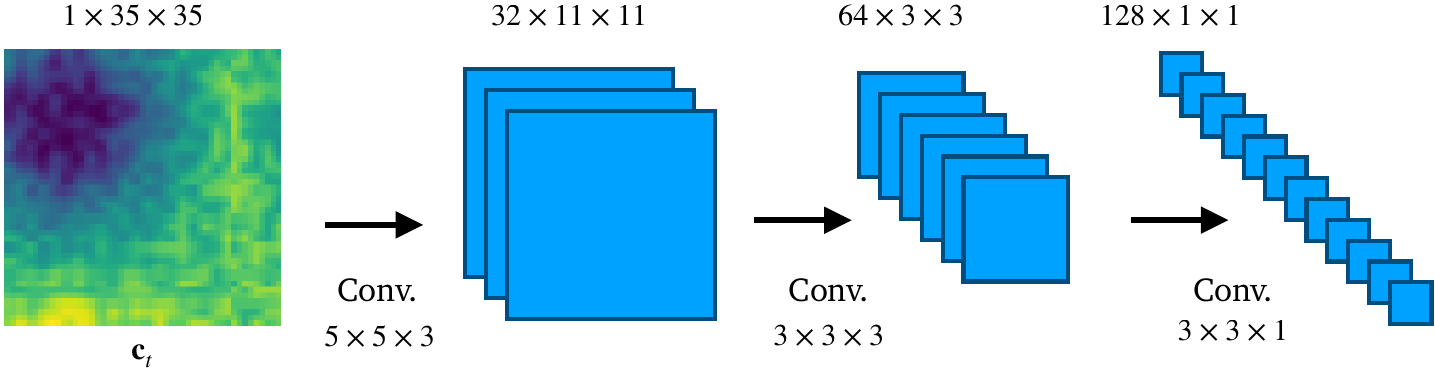}\vspace{-8pt}
    \caption{Network structure of the heightmap encoder $\mathcal{G}_\phi$ giving additional control input $\mathbf{c}_t$. The convolution operation \texttt{Conv.} has format of kernel height $\times$ kernel width $\times$ stride. The dimension of $\mathbf{c}_t$ and the feature maps are shown in the format of channels $\times$ height $\times$ width.}\vspace{-8pt}
    \label{fig:cnn}
\end{figure}

\begin{figure}[t]
    \centering
    \includegraphics[width=.496\linewidth]{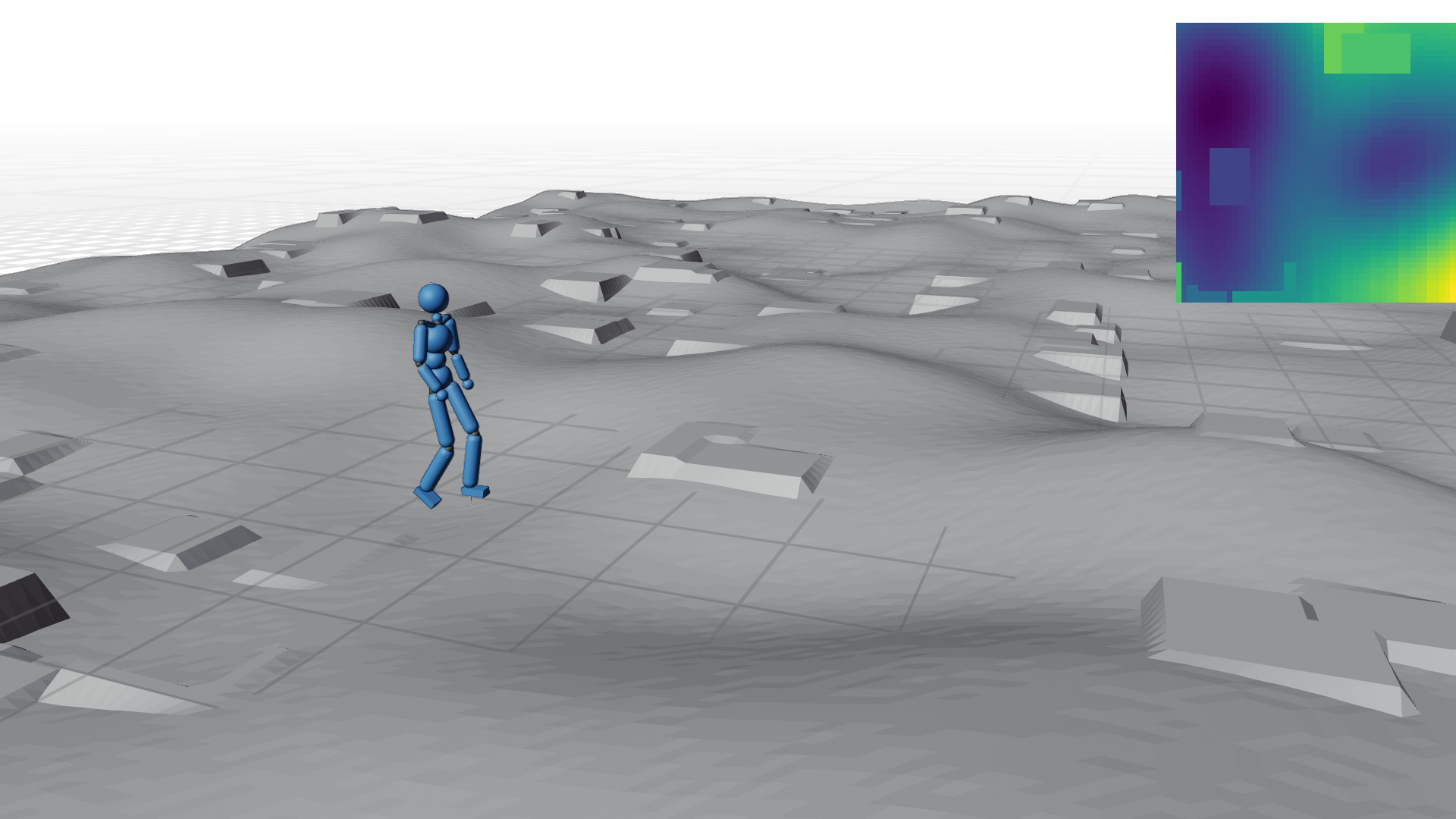}\hfill
    \includegraphics[width=.496\linewidth]{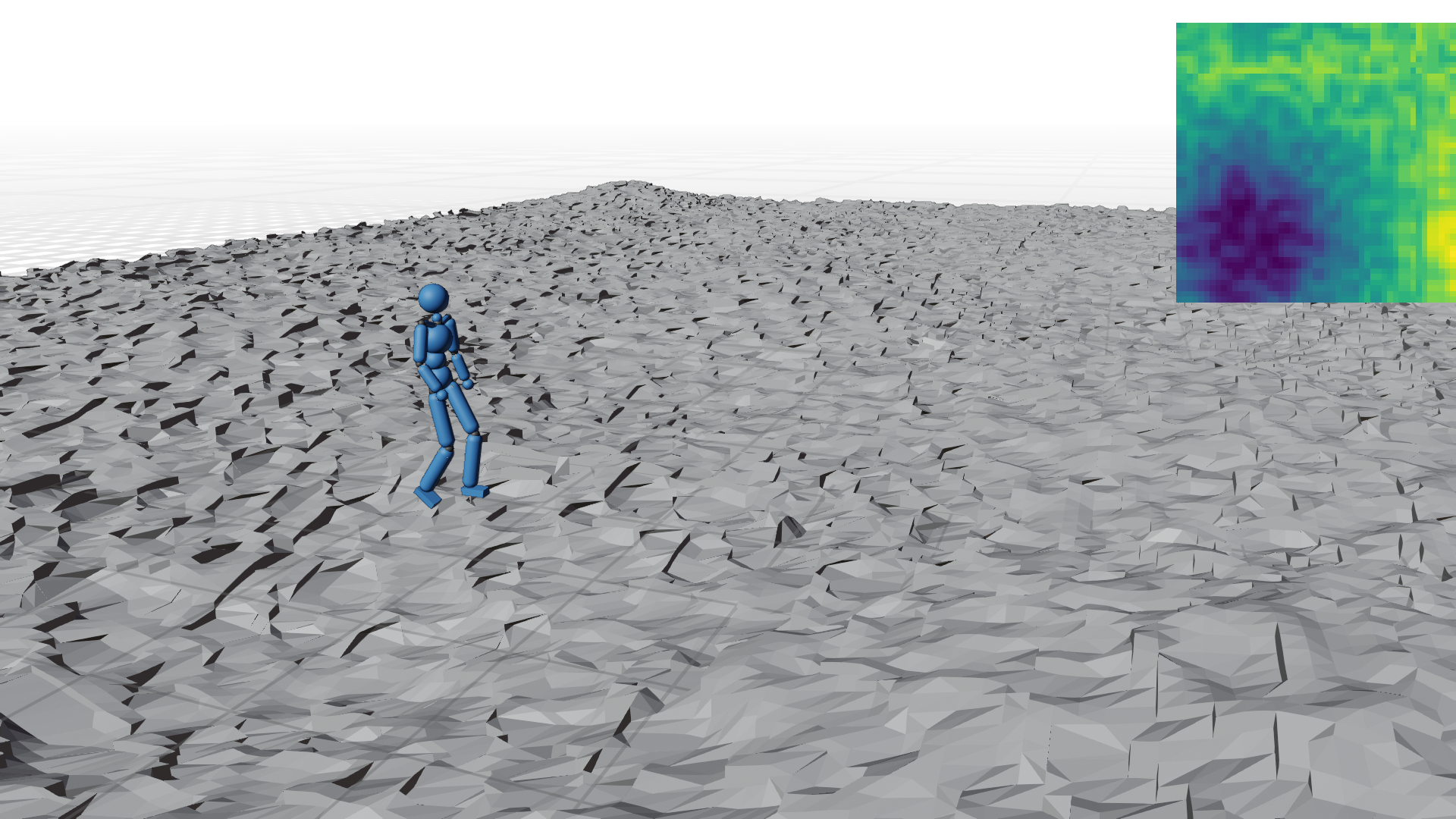}\vspace{-8pt}
    \caption{Procedurally generated terrains with minimaps showing the local heightmaps used as additional control input $\mathbf{c}_t$. The $35\times35$ heightmaps encode a 3.4~m$\times$3.4~m area centered around the character.
    }\vspace{-10pt}
    \label{fig:terrain_exp}
\end{figure}

The hyperparameters used for policy training are listed in Table~\ref{tab:hyper}.
Half of the training samples used by the discriminator are drawn from the simulated character and half are sampled from the reference motions.
The learning objective weight $\omega_k$ in Equation~9 of the main text %
is 0.5 for imitation and 0.5 for the task of goal-steering navigation.
During adaptation, we use 0.35 for imitation and 0.65 for navigation in
motion style transfer tasks and 0.5 for both imitation and navigation in other tasks.

\section{Terrain Maps}
Terrains are generated using a combination of Perlin noise 
and procedural generation.
To generate more challenging terrains,
we add extra uniform noise to make the terrain more uneven, while including some randomly generated rectangular blocks for more stepped terrain. 
In the most rugged
example, we generate terrain by applying uniform noise on a terrain composed of multiple surfaces with slopes varying from -0.3 to 0.3. An extra slope threshold is applied to cut off the bulges on the terrain and make it rugged.
The generated terrains in all %
examples are normalized to have a maximal height of 0.75~m and a minimum height of -0.75~m.
Figure~\ref{fig:cnn} shows the network structure of the heightmap encoding module $\mathcal{G}_\phi$, which has three convolutional layers.
Following previous work~\cite{mnih2013playing}, we do not use any pooling layer in the network.
Resulting terrain examples appear in Figure~\ref{fig:terrain_exp}.

\end{document}